\PassOptionsToPackage{table,rgb}{xcolor}
\documentclass[sigconf,screen]{acmart}

\usepackage{enumitem}
\usepackage{amsmath,amsfonts}
\usepackage{algorithmic}
\usepackage{array}
\usepackage[caption=false,font=normalsize,labelfont=sf,textfont=sf]{subfig}
\usepackage{textcomp}
\usepackage{stfloats}
\usepackage{url}
\usepackage{verbatim}
\usepackage{graphicx}
\usepackage{balance}
\usepackage{adjustbox}
\usepackage{booktabs}
\usepackage{multirow}
\usepackage{threeparttable}
\usepackage{arydshln}
\usepackage{tcolorbox}
\usepackage{soul}
\usepackage{fontawesome}
\usepackage{makecell}
\usepackage{natbib}
\usepackage{colortbl}
\usepackage{siunitx}
\usepackage{microtype}
\definecolor{mygray}{gray}{.9}
\usepackage{tikz}
\usepackage[T1]{fontenc}
\usepackage{aecompl}

\usepackage{datenumber}

\definecolor{chestnut}{rgb}{0.8, 0.36, 0.36}

\definecolor{chestnut}{rgb}{0.8, 0.36, 0.36}

\newcounter{dateone}\newcounter{datetwo}%
\newcommand{\daydifftoday}[3]{%
\setmydatenumber{dateone}{\the\year}{\the\month}{\the\day}%
\setmydatenumber{datetwo}{#1}{#2}{#3}%
\addtocounter{datetwo}{-\thedateone}%
\thedatetwo
}

\AtBeginDocument{
  \providecommand\BibTeX{{
    \normalfont B\kern-0.5em{\scshape i\kern-0.25em b}\kern-0.8em\TeX}}}

\newcommand{\Comment}[1]{}

\setcopyright{acmlicensed}
\acmDOI{10.1145/3650212.3680395}
\acmYear{2024}
\copyrightyear{2024}
\acmISBN{979-8-4007-0612-7/24/09}
\acmConference[ISSTA '24]{Proceedings of the 33rd ACM SIGSOFT International Symposium on Software Testing and Analysis}{September 16--20, 2024}{Vienna, Austria}
\acmBooktitle{Proceedings of the 33rd ACM SIGSOFT International Symposium on Software Testing and Analysis (ISSTA '24), September 16--20, 2024, Vienna, Austria}
\begin{document}

% \title{Deep Learning Program Semantics to Detect Equivalent Mutants}
% \title{An Empirical Study of Large Language Models for Equivalent Mutant Detection}
\title{Large Language Models for Equivalent Mutant Detection: How Far Are We?}

\author{Zhao Tian}
\orcid{0000-0002-9316-7250}
\affiliation{
  \institution{College of Intelligence and \\
  Computing, Tianjin University}
  \city{Tianjin}
  \country{China}
}
\email{tianzhao@tju.edu.cn}

\author{Honglin Shu}
\orcid{0009-0005-7311-7060}
\affiliation{
  \institution{Kyushu University}
  \city{Fukuoka}
  \country{Japan}
}
\email{shu.honglin.167@s.kyushu-u.ac.jp}

\author{Dong Wang}
\authornotemark[1]
% \authornote{Dong Wang is the corresponding author.}
\orcid{0000-0002-2004-0902}
\affiliation{
  \institution{College of Intelligence and \\
  Computing, Tianjin University}
  \city{Tianjin}
  \country{China}
}
\email{dong_w@tju.edu.cn}

\author{Xuejie Cao}
\orcid{0009-0003-6194-8110}
\affiliation{
  \institution{College of Intelligence and \\
  Computing, Tianjin University}
  \city{Tianjin}
  \country{China}
}
\email{caoxuejie@tju.edu.cn}

\author{Yasutaka Kamei}
\orcid{0000-0002-7058-1045}
\affiliation{
  \institution{Kyushu University}
  \city{Fukuoka}
  \country{Japan}
}
\email{kamei@ait.kyushu-u.ac.jp}

\author{Junjie Chen}
\authornotemark[1]
% \authornote{Junjie Chen is the corresponding author.}
\orcid{0000-0003-3056-9962}
\affiliation{
  \institution{College of Intelligence and \\
  Computing, Tianjin University}
  \city{Tianjin}
  \country{China}
}
\email{junjiechen@tju.edu.cn}

% \authornote{Dong Wang and Junjie Chen are the corresponding authors.}
\thanks{*Dong Wang and Junjie Chen are the corresponding authors.}

\begin{abstract}
Mutation testing is vital for ensuring software quality. 
However, the presence of equivalent mutants is known to introduce redundant cost and bias issues, hindering the effectiveness of mutation testing in practical use. 
Although numerous equivalent mutant detection (EMD) techniques have been proposed, they exhibit limitations due to the scarcity of training data and challenges in generalizing to unseen mutants. 
Recently, large language models (LLMs) have been extensively adopted in various code-related tasks and have shown superior performance by more accurately capturing program semantics. 
Yet the performance of LLMs in equivalent mutant detection remains largely unclear. 
In this paper, we conduct an empirical study on 3,302 method-level Java mutant pairs to comprehensively investigate the effectiveness and efficiency of LLMs for equivalent mutant detection.
Specifically, we assess the performance of LLMs compared to existing EMD techniques, examine the various strategies of LLMs, evaluate the orthogonality between EMD techniques, and measure the time overhead of training and inference.
Our findings demonstrate that LLM-based techniques significantly outperform existing techniques (i.e., the average improvement of $35.69\%$ in terms of F1-score), with the fine-tuned code embedding strategy being the most effective. 
Moreover, LLM-based techniques offer an excellent balance between cost (relatively low training and inference time) and effectiveness.
Based on our findings, we further discuss the impact of model size and embedding quality, and provide several promising directions for future research.
This work is the first to examine LLMs in equivalent mutant detection, affirming their effectiveness and efficiency.
\end{abstract}




\maketitle

% \vspace{-2mm}
\section{Introduction}
\label{sec:introduction}
% \wang{Comment}
% \tian{Comment}
% \shu{Comment} 
% \chen{Comment} 
% \kevin{Comment} 
% \yasu{Comment}
% Basic Intro Flow:
% (Concept of Mutation Testing and Mutant) + (Concept of equivalent mutant + its impact) + (Automatic identification method + Limitations) + (LLM popularity and its excellent capability of capturing program semantics and have shown promise in several downstream code tasks) + (Little is known about the LLM on such significant software testing, thus we conduct an empirical study to blablabla)

Mutation testing~\cite{andrews2005mutation, jia2010analysis} involves injecting a set of intentional artificial faults into a program being tested, to measure the effectiveness of a test suite (and further enhance it).
In this context, a program with an artificial fault is referred to as a \textit{mutant}, constructed by deliberately changing a small portion of code in the program under test.
Besides measuring test effectiveness, mutation testing has been extensively extended to facilitate other software testing and debugging tasks (e.g.,
test case prioritization~\cite{lou2015mutation}, bug detection~\cite{pradel2018deepbugs}, and fault localization~\cite{papadakis2015metallaxis}) achieving state-of-the-art performance.

Despite its popularity and importance, mutation testing is still plagued by high costs, notably exacerbated by the presence of equivalent mutants~\cite{papadakis2019mutation,titcheu2020selecting}, a problem known to be undecidable for over three decades.
An equivalent mutant is redundant because it exhibits the same behavior as the original program for all possible test cases~\cite{jia2010analysis,madeyski2013overcoming}.
Recent research indicates that the rate of equivalent mutants in real-world development scenarios ranges from 4\% to 39\%~\cite{madeyski2013overcoming}.
Moreover, equivalent mutants introduce significant bias into mutation-based analysis.
Specifically, the widely-used metric, the \textit{mutation score}, is calculated using only non-equivalent mutants. Therefore, the presence of equivalent mutants makes it impossible to achieve a score of 100 percent.
As a result, developers might not fully trust an otherwise sufficient test suite.
%The cost and bias issues associated with equivalent mutants can impede the practical use of mutation testing, even slow down the software development process, and negatively impact software quality~\cite{grun2009impact}. 
The cost and bias issues associated with equivalent mutants can impact the practical effectiveness of mutation testing, potentially slowing down the software development process and negatively affecting software quality.
Hence, detecting redundant equivalent mutants has become increasingly critical.

Over the years, numerous equivalent mutant detection (EMD) techniques have been proposed to tackle the equivalent mutant problem~\cite{kintis2017detecting,gheyi2021identifying}.
% Earlier traditional EMD techniques encompassed genetic algorithms~\cite{adamopoulos2004overcome}, constraint-based testing~\cite{offutt1997automatically,kushigian2019medusa,baer2020mutantdistiller}, coverage analysis~\cite{schuler2010covering,papadakis2013mutation,papadakis2014mitigating}, automata language equivalence~\cite{devroey2018model}, software behavior graph~\cite{gong2022equivalent}, and compiler optimization~\cite{papadakis2015trivial,houshmand2017tce+,kintis2017detecting}.
Traditional EMD techniques often rely on pre-defined rules such as constraint-based testing~\cite{offutt1997automatically,schuler2013covering,kushigian2019medusa,baer2020mutantdistiller} and compiler optimization~\cite{papadakis2015trivial,houshmand2017tce+,kintis2017detecting}, showing limited performance in complex practical development scenarios~\cite{naeem2020machine,peacock2021automatic}.
Meanwhile, more advanced learning-based EMD techniques have been proposed, including conventional machine learning-based classifiers (e.g., KNN and SVM)~\cite{naeem2020machine,brito2020preliminary,chung2022augmenting} and tree-based neural network models~\cite{peacock2021automatic}. 
While these learning-based EMD techniques improve upon traditional techniques by comparing extracted code features, they may not fully capture program semantics, especially when it comes to minor syntax differences.
Additionally, their effectiveness is limited by the scarcity of training data for equivalent mutants and potential challenges in generalizing to unseen mutants.

Lately, large language models (LLMs) have demonstrated impressive performance in both natural language processing (NLP)~\cite{touvron2023llama,song2023llm} and software engineering (SE)~\cite{schafer2023empirical,sallou2024breaking}.
Furthermore, given that the pre-trained corpus of these LLMs (e.g., StarCoder~\cite{li2023starcoder} and Code Llama~\cite{touvron2023llama}) contains a vast amount of code snippets, they can learn generalized knowledge, thereby, in turn, boosting a variety of code-related tasks~\cite{yang2024empirical,liu2024your}.
Particularly, LLMs have shown promise in diverse software testing scenarios by using different learning strategies like fine-tuning and prompt engineering for tasks such as test case generation~\cite{schafer2023empirical, yang2024enhancing}, program debugging~\cite{feng2024prompting, li2023nuances}, and program repair~\cite{huang2023empirical}.  
% For example, \citet{schafer2023empirical} proposed an test generation technique, employing prompt engineering to automatically generating high-quality unit tests.
% \citet{huang2023empirical} conducted a comprehensive study on the repair capability of five popular LLMs with the fine-tuning paradigm, suggesting that the LLM-based techniques can significantly outperform previous state-of-the-art Automated Program Repair (APR) techniques.
Detecting equivalent mutants is closely associated with understanding code semantics. 
LLMs, pre-trained on extensive code snippets from diverse resources, possess a superior grasp of code semantics compared to the above traditional learning-based EMD techniques that lack sufficient pre-training. 
Consequently, LLMs are more likely to effectively address the issue of data scarcity.
However, there is a lack of comprehensive understanding of how well LLMs perform in detecting equivalent mutants.
Encouraged by the remarkable performance of LLMs, we conjecture that they can better understand and distinguish code semantics between equivalent mutants, even with minor syntax differences.

In our paper, we conduct an empirical study on 3,302 method-level Java
mutant pairs to delve into the potential of leveraging LLMs for detecting equivalent mutants.
To comprehensively assess the effectiveness and efficiency of LLM-based techniques, we formulate the following four research questions with their motivations:\\
\noindent
    \textbf{RQ1: \ul{What is the performance of state-of-the-art LLMs on equivalent mutant detection?}}
    We first aim to explore the capability of LLMs in detecting equivalent mutants. 
    Specifically, we compare LLMs with ten typical or state-of-the-art existing EMD techniques to determine whether LLMs are superior.
    \\
\noindent
\textbf{Results:} LLM-based techniques significantly surpass all ten EMD baselines in detecting equivalent mutants, with average F1-score improvements of 75.18\% for Compiler-based techniques, 19.14\% for ML-based techniques, and 12.75\% for Tree-based NN techniques.
\\ 
    \noindent
    \textbf{RQ2: \ul{What is the best strategy using LLMs for equivalent mutant detection?}}
    Different strategies (e.g., code embedding and prompting) utilized by LLMs may influence the detection performance.
    Thus, we further investigate the extent of their impact, which could offer insights into the optimal selection of strategies for enhancing LLM performance on equivalent mutant detection.\\
\noindent
\textbf{Results:} 
The fine-tuned code embedding strategy demonstrates the superiority of equivalent mutant detection.
Specifically, the fine-tuned UniXCoder outperforms all the combinations of LLMs and strategies with the improvement of 1.16\%$\sim$78.85\% in terms of F1-score.
On the other hand, LLMs based solely on prompting strategies cannot achieve comparable performance.
\\ 
    \noindent
    \textbf{RQ3: \ul{What degree of orthogonality exists between our studied EMD techniques?}}
    % Certain types of equivalent mutant detection techniques may be prone to identify mutant operators.
    Certain EMD categories and LLM strategies may be prone to identify mutants based on specific mutation operators.
    Hence, RQ3 seeks to gain an understanding of the characteristics of various EMD categories and LLM strategies by analyzing the orthogonality between them.
    \\
\noindent
\textbf{Results:} The LLM-based techniques and the fine-tuned code embedding strategy significantly surpass the other EMD categories and LLM strategies in terms of the unique correct/incorrect detections and the detection performance on each mutation operator, reinforcing the findings of RQ1 and RQ2. 
\\ 
    \noindent
    \textbf{RQ4: \ul{How efficient are our studied EMD techniques?}}
    As the size of LLMs has increased exponentially recently, the cost of applying these large models has also grown significantly.
    While larger LLMs may perform better, the trade-off between their detection performance and the cost is crucial.
    In this RQ, we comprehensively investigate the time efficiency (i.e., training and inference time) of all studied EMD techniques.\\
    \noindent
    \textbf{Results:} The inference time of the best-performing LLM-based technique (\SI{0.0431}{s}) exceeds that of the best-performing Compiler-based technique (\SI{2.3537}{s}), but is marginally longer than that of the best-performing ML-based technique (\SI{0.0019}{s}) and the best-performing Tree-based NN technique (\SI{0.0274}{s}). 
    The results highlight the LLM's excellent balance between cost and effectiveness.
\\
    \noindent
    \textbf{Contributions.}
To sum up, the contributions of this study are:
\begin{itemize}[leftmargin=10pt]
    \item We perform the first large-scale empirical study to assess the capability of LLMs for equivalent mutant detection, considering four perspectives (i.e., effectiveness compared to existing EMD techniques, the impact of LLM strategies, orthogonality between various EMD techniques, and time efficiency).

    \item The study confirms the superiority of LLM-based equivalent mutant detection techniques, yielding state-of-the-art performance.
    
    \item We provide valuable insights into the capabilities and limitations of LLMs for equivalent mutant detection. 
    The findings will serve as essential guidance for future research aimed at enhancing LLM-based equivalent mutant detection and other aspects of software engineering.
    Additionally, we open source all data, code, and analysis details involved in our study~\cite{EMD2024}.
\end{itemize}

% \noindent
% \textbf{Paper Organization.}
% The remainder of the paper is organized as follows. 
% Section~\ref{sec:background} explains the background. 
% Section~\ref{sec:evaluation_design} introduces our study design. 
% Section~\ref{sec:results_and_analysis} reports the experimental results and analysis. 
% Section~\ref{sec:discussion} discusses the lessons learnt and future work. 
% Section~\ref{sec:conclusion} concludes our study.
% Section~\ref{sec:data_availability} provide a supporting statement on the data availability.
\section{Background and Related Work}
\label{sec:background}
% \begin{itemize}
%     \item Trivial compiler equivalence: A large scale empirical study of a simple, fast and effective equivalent mutant detection technique - 2015
%     \item Detecting trivial mutant equivalences via compiler optimisations - TSE 2017
%     \item A systematic literature review of techniques and metrics to reduce the cost of mutation testing - JSS 2019
% \end{itemize}

% \subsection{Mutation Testing}
% \label{subsec:mutation_testing}

\subsection{Mutation Testing}
\label{subsec:equivalent_mutants}
Mutation testing is a program analysis approach that involves artificially altering the source code to inject (likely) faulty behavior~\cite{shi2019mitigating,moradi2023muakka}.
The changing rules are called \textit{mutation operators}, which are typically constructed based on syntactic rules derived from the grammar of the target programming language~\cite{ojdanic2023mutation}.
% , i.e., replacing the relational operator \textbf{\texttt{>}} with \textbf{$<$}.
For instance, by applying the relational operator replacement mutation operator, the code fragment "\texttt{if(x==y)}" in the original program can be changed to "\texttt{if(x!=y)}", thereby constructing a mutant.
Its basic assumption is that the introduced faults can effectively represent real faults~\cite{andrews2006using,gopinath2014mutations}.
Mutation testing aims to evaluate the effectiveness of the test suits~\cite{gopinath2018if,perretta2022use}.
A mutant is referred to be "killed" if it is
detected by any test case; otherwise, it is said to be "live".
% The key metric of mutation testing is the mutation score, also known as the percentage of killed mutants. This score is calculated by dividing the number of killed mutants, those which cause test cases to fail, by the total number of generated mutants.
The key metric of mutation testing is the mutation score, also known as the percentage of killed mutants. This score is calculated by dividing the number of killed mutants, those that cause test cases to fail, by the total number of generated non-equivalent mutants.

% \vspace{-2mm}
\subsection{Equivalent Mutants}

The equivalent mutant problem (EMP) is a critical issue in mutation testing that has been extensively studied for decades~\cite{kintis2017detecting,bartocci2023property}.
A mutant is deemed equivalent if, for all possible test cases, it exhibits the same behavior as the original program under test.  
These mutants are syntactically different but semantically equivalent to the original program, and cannot be killed by any test cases.
% Equivalent mutants are considered one of the main causes why mutation testing is seldom used in practice as they consume resources without producing any useful test~\cite{yao2014study}.
% Since the mutation score is calculated only using non-equivalent mutants and does not encompass the complete detection of all equivalent mutants, reaching a mutation score of 100 percent is impossible.
% Consequently, programmers may lack complete confidence in the adequacy of a potentially perfectly adequate test set.
Equivalent mutants are often considered one of the main causes contributing to the limited adoption of mutation testing in practice due to their high computational cost and introduction of significant bias~\cite{yao2014study,papadakis2019mutation}.
Prior research has found that the rate of equivalent mutants in real-world development scenarios might lie between 4\% and 39\%~\cite{madeyski2013overcoming}.
The generation of a high number of mutants leads to increased computational costs and bias for their evaluation~\cite{holling2016nequivack}, and a significant effort is required to detect equivalent mutants.

Detecting equivalent mutants in practice is challenging since in code mutation, program equivalence is undecidable~\cite{arcaini2017novel,kim2022predictive}.
% Automatically detecting all equivalent mutants in practice is challenging, since program equivalence is undecidable in mutation testing~\cite{budd1982two}.
% A series of approaches have been developed to address this issue.
A series of EMD techniques have been developed to address this issue.
In earlier times, methods such as genetic algorithms~\cite{adamopoulos2004overcome}, constraint-based testing~\cite{offutt1997automatically,kushigian2019medusa,baer2020mutantdistiller}, coverage analysis~\cite{schuler2010covering,schuler2013covering,papadakis2013mutation,papadakis2014mitigating}, automata language equivalence~\cite{devroey2018model}, software behavior graphs~\cite{gong2022equivalent}, dynamic subsumption relations~\cite{guimaraes2020optimizing,gheyi2021identifying}, and compiler optimization~\cite{papadakis2015trivial,houshmand2017tce+,kintis2017detecting} were used to identify equivalent mutants.
Recently, learning-based EMD techniques have been proposed, including conventional binary classifiers (e.g., KNN and SVM)~\cite{naeem2020machine,brito2020preliminary,chung2022augmenting} and tree-based neural network (NN) models~\cite{peacock2021automatic}. 
In particular, an early assessment of the tree-based NN technique with 582 mutants, derived from only two mutation operators, yielded promising results with a classification accuracy of 90\%~\cite{peacock2021automatic}.

Among these existing EMD techniques, certain techniques involve feature extraction through the execution of mutant programs within the context of the test suite~\cite{baer2020mutantdistiller,naeem2020machine,chung2022augmenting,gong2022equivalent}. 
Although leveraging information from test suites can enhance the performance of equivalent mutant detection, generating and executing a large number of test cases in practice consumes significant time and computational resources~\cite{jia2010analysis}. 
Therefore, in this paper, we only study those techniques focusing on the semantic learning of code by examining the capability of LLMs for equivalent mutant detection.
% In selecting baselines, we conducted a concise literature review and snowballing, opting for three typical and state-of-the-art baselines (as introduced in Section~\ref{subsec:baselines}) developed over the past decade that do not rely on test suite execution information.

% \begin{tcolorbox}[colback=gray!5]
% \textbf{Motivation I}:
% \end{tcolorbox}

\subsection{Large Language Models}
\label{sebsec:llms}
% Pre-trained language models (PLMs) have demonstrated impressive capabilities in solving various NLP tasks.
% In recent years, researchers have observed that scaling up the model sizes
% significantly enhances their capacity, leading to remarkable performance improvements when the
% parameter scale surpasses a certain threshold 

Large language models have become a dominant part of NLP because of their exceptional performance, such as Llama 2~\cite{touvron2023llama} and PaLM 2~\cite{anil2023palm}.
% Due to Scaling Laws, the size of LLMs has increased significantly in recent years. 
% For instance, StarCoder~\cite{li2023starcoder} and Llama 2~\cite{Touvron2023Llama2O} have over billion parameter LLMs within large-scale Transformer architecture~\cite{vaswani2017attention}. 
% GPT-4~\cite{openai2023gpt4} and Llama 2~\cite{Touvron2023Llama2O} have over one hundred billion parameters within large-scale Transformer architecture~\cite{vaswani2017attention}.
Aside from the general purposes of LLMs, many LLMs have been trained on code corpora for transferring the impressive text generation capability to the code domain, such as StarCoder~\cite{li2023starcoder} and Code Llama~\cite{roziere2023code}.

Software testing with LLMs recently has undergone significant growth~\cite{Wang2023SoftwareTW}.
Particularly, LLMs are widely used for test case generation~\cite{schafer2023empirical,xia2023universal}, program debugging~\cite{feng2024prompting,li2023nuances}, and program repair~\cite{huang2023empirical} through various learning strategies such as fine-tuning and prompt engineering.
To name a few, in terms of test case generation, \citet{schafer2023empirical} presented a large-scale empirical evaluation of the effectiveness of LLMs for automatic unit test generation with prompting strategies.
\citet{xia2023universal} proposed Fuzz4All, which leverages LLMs as the mutation engine to produce diverse and realistic inputs for any practically relevant language, outperforming the existing language-specific fuzzers.
In terms of program debugging, \citet{feng2024prompting} introduced a lightweight approach namely AdbGPT to reproduce the bugs from bug reports through prompt engineering.
\citet{li2023nuances} developed a technique, Differential Prompting, to effectively find failure-inducing test cases with the help of the compilable code synthesized by the inferred intention.
In terms of program repair, \citet{huang2023empirical} conducted a comprehensive study on the repair capability of five popular LLMs with the fine-tuning paradigm, suggesting that the LLM-based methods can significantly outperform previous APR techniques.

Despite these attempts, the effectiveness of LLMs in equivalent mutant detection remains largely unexplored. 
Recently, ~\citet{ma2023scope} conducted a preliminary study on equivalent mutant detection using a small dataset (i.e., 200 mutants) with ChatGPT. 
To address the limitations of dataset size and improve generalizability, our study extensively evaluates the performance of typical and state-of-the-art LLMs in equivalent mutant detection considering diverse aspects including strategies, orthogonality, and efficiency.

% Furthermore, LLM are not only employ in three aforementioned testing-related generation tasks, but also largely applied in bug analysis~\cite{Feng2023PromptingIA}, debug~\cite{Yang2023LargeLM}, and program repair~\cite{Gao2023WhatMG}. 
% Despite LLMs being applied in various software testing tasks, the application of LLMs in mutant testing, particularly equivalent mutant detection, remains under-explored.
% \wang{Please highlight that Ma's Chatgpt work to motivate.}

% \begin{tcolorbox}[colback=gray!5]
% \textbf{Motivation II}:
% \end{tcolorbox}
\section{Study Design}
\label{sec:evaluation_design}

\begin{figure}[t!]
    \centering
    % \vspace{-.2cm}
    \includegraphics[width=1.0\linewidth]{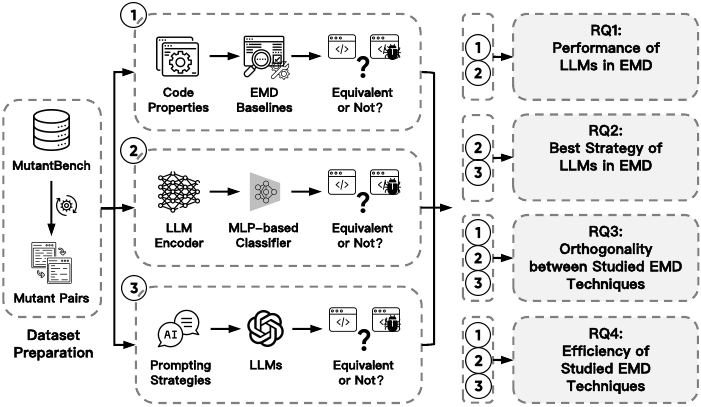}
    \caption{Overview of experimental design. \textcircled{1}/\textcircled{2}/\textcircled{3} represents the workflow of EMD baselines/code embedding strategies/prompting strategies, respectively.}
    \label{fig:overview}
    % \vspace{-.5cm}
    % \vspace{-4mm}
\end{figure}

Figure~\ref{fig:overview} illustrates the overview of our study design.
Initially, based on the most widely-used MutantBench~\cite{van2021mutantbench} dataset, we construct the training and test datasets comprising code pairs of the original program and its mutant.
We first explore the effectiveness of the existing EMD techniques and state-of-the-art LLMs on equivalent mutant detection (RQ1).
We also investigate the extent of impact resulting from various strategies utilized by LLMs on equivalent mutant detection (RQ2). 
Then, we gain an understanding of the characteristics of various EMD techniques by analyzing the orthogonality between them (RQ3).
Finally, we quantitatively measure the time efficiency across the studied EMD techniques (RQ4).

\begin{table}[t]
    \caption{Statistics of Java programs from MutantBench}
    \label{tab:dataset}
    \centering
    \tabcolsep=2.2mm
    % \vspace{-3mm}
    \small
    \begin{adjustbox}{max width=1.0 \textwidth,center}
        \begin{tabular}{ lcccc }
            \toprule
            \toprule
            \textbf{Dataset} & \textbf{\#Programs} & \textbf{\#Methods} & \textbf{\#EQ} & \textbf{\#NEQ} \\ 
        	\midrule
            \textit{$MutantBench_{train}$} & \multirow{2}{*}{19} & \multirow{2}{*}{328} & 250 & 1,402 \\
            \textit{$MutantBench_{test}$}  &  & & 249 & 1,401 \\
            \bottomrule
            \bottomrule
        \end{tabular}
    \end{adjustbox}
    \begin{tablenotes}
        \footnotesize
        \item[*] \#Programs, \#Methods, \#EQ, and \#NEQ represent the number of programs, methods, equivalent mutants, and non-equivalent mutants, respectively.
    \end{tablenotes}
    % \vspace{-5mm}
\end{table}

% \vspace{-2mm}
\subsection{Dataset Preparation}
\label{subsec:dataset}

\

\textbf{Studied dataset}. 
% original sentences
% To evaluate the performance of equivalent mutant detection, we choose the most widely-used MutantBench~\cite{van2021mutantbench}.
% The dataset comprises 37 programs written in both C/C++ and Java programming languages, with a total of 4,400 mutants, of which 32.27\% are equivalent mutants.
To evaluate the performance of EMD techniques, we select the most widely-used MutantBench~\cite{van2021mutantbench}, containing 4,400 mutant pairs in both C/C++ and Java programming languages.
More specifically, MutantBench consolidates many existing open-source datasets~\cite{yao2014study,kintis2016analysing, houshmand2017tce+, baer2020mutantdistiller} into one benchmark, enhancing dataset diversity and encompassing a broader spectrum of mutant types.
In this study, we focus on the primary programming language, Java.
Specifically, it accounts for 3,302 mutant pairs from 19 programs, as depicted in Table~\ref{tab:dataset}.
% \shl{3,302 mutant pairs or mutants?}
% The dataset comprises 37 programs in both C and Java, totaling 4400 mutants, among which 32.27\% are equivalent mutants.
% Following existing work~\cite{chung2022augmenting}, we use all 19 Java programs from MutantBench (with a total of 3302 mutants), whose details are shown in the Table~\ref{tab:dataset}. 

% \tz{Please describe the process of data processing. Note that we extract method-level code pairs}

% \noindent 
\textbf{Data pre-processing}. 
We carried out data pre-processing in two steps to meet the input format requirements of LLMs.
Due to the constraint imposed by the maximum input token length of LLMs, we first removed all the natural language comments from the Java code utilizing pre-defined regular expressions.
It is noted that natural language comments do not contribute to the equivalent mutant detection task.
Second, we opted to detect the equivalence of mutant pairs at the method level instead of the program level, thereby reducing the input length.
Since all mutation operators in the MutantBench dataset are applied exclusively within a single method rather than across multiple methods, ensuring that method-level mutant pairs preserve the semantic equivalence of the original mutant pairs.
Hence, following existing work~\cite{tufano2019learning,tian2022learning}, we also separated the original program and its corresponding mutant into method-level pieces.
We then identified and selected the methods containing the mutation location from both the original program and its mutant.
This resulted in 3,302 method-level Java mutant pairs, all of which are devoid of natural language comments.

% \begin{itemize}
%     \item Filtering out all comments in the Java code to make the dataset more cohesive. \wang{using regular expression? need some details}
%     \item Extracting method-level mutant pairs from the original program and corresponding mutant programs by using a common practice~\cite{tufano2019learning,tian2022learning}. \wang{what is common practice? any detailed method? one sentence to describe.}
% \end{itemize} 

% \noindent 
\textbf{Construction of training and test datasets}.
Following existing work~\cite{cervellera2017distribution,lu2021variance}, we adopted a stratified sampling strategy to reduce sampling bias and ensure that both training and test datasets are representative of the entire dataset. 
Firstly, the 3,302 total mutant pairs were divided into two subgroups: $MutantBench_{eq}$, comprising solely equivalent mutant pairs, and $MutantBench_{neq}$, encompassing the remaining non-equivalent mutant pairs.
Subsequently, $\sim$50\% of the mutant pairs from both $MutantBench_{eq}$ and $MutantBench_{neq}$ were randomly selected to construct a training dataset, denoted as $MutantBench_{train}$. 
Similarly, the remaining mutant pairs from both subgroups were amalgamated to form a test dataset, named $MutantBench_{test}$.
Finally, we obtained 1,652 mutant pairs in $MutantBench_{train}$, comprising 250 equivalent mutant pairs and 1,402 non-equivalent mutant pairs. 
For $MutantBench_{test}$, we ended up with 1,650 mutant pairs, comprising 249 equivalent mutant pairs and 1,401 non-equivalent mutant pairs.
In particular, we confirmed that there is no data leakage between our training and test datasets through manual inspection.

% \vspace{-3mm}
\subsection{Experimented Large Language Models}
\label{subsec:models}

In our study, we investigated the performance of ten state-of-the-art LLMs for equivalent mutant detection. 
These models have been widely adopted in the literature ~\cite{du2023pre, tian2023code, Wang2023SoftwareTW}, including:
\begin{itemize}[leftmargin=10pt]
    \item \textbf{CodeBERT}~\cite{feng2020codebert} is a popular pre-trained model designed to learn from bimodal data encompassing both source code and natural languages, using a multilayer Transformer architecture.
    \item \textbf{GraphCodeBERT}~\cite{guo2020graphcodebert} is a pre-trained model that leverages semantic-level code information to enhance code representation using a transformer-based architecture.
    % It significantly improves code understanding tasks, including code clone detection and code search, based on the code structure information~\cite{khajezade2024investigating}.

    \item \textbf{PLBART}~\cite{ahmad2021unified} is a bidirectional and autoregressive transformer, which adopts a BART architecture and employs denoising objectives for pre-training on unlabeled data spanning source code and natural language. 
    % It learns multilingual representations suitable for various program understanding and generation tasks.

    \item \textbf{CodeT5}~\cite{wang2021codet5} is a unified encoder-decoder model that incorporates token type information in code. 
    It extends the T5 architecture, utilizing denoising sequence-to-sequence pre-training. 

    \item \textbf{CodeT5+}~\cite{wang2023codet5+} builds upon CodeT5.
    It uses a shallow encoder and a deep decoder, and is trained in multiple stages, initially with unimodal data, and later with bimodal data.
    % The architectural enhancements and advanced training strategies enable CodeT5+ to learn rich representations.
    % (e.g., span denoising and contrastive learning) 

    \item \textbf{UniXCoder}~\cite{guo2022unixcoder} is a unified cross-modal pre-trained model that exploits multi-modal information, such as abstract syntax tree (AST) and code comments, to improve code representation.
    It is also based on transformer-based architecture.
    % and excels in code comprehension and code generation tasks~\cite{wang2023you}.

    % , thereby achieving state-of-the-art performance in a variety of code understanding and generation tasks~\cite{liu2024reliability}.
    
    \item \textbf{StarCoder}~\cite{li2023starcoder} is based on SantaCoder architecture. It possesses its own encoder model, StarEncoder, and features infilling capabilities and rapid, large-batch inference made possible by multi-query attention.
    % , that can be efficiently applied to diverse code-related and text-related tasks (e.g., code generation and masked language modeling).
    
    \item \textbf{Code Llama}~\cite{roziere2023code} is one of the most popular LLMs for code generation and infilling derived from Llama 2 models.
    It is a decoder-only model and additionally fine-tuned on 500B tokens from an extra code-heavy dataset.
    % According to previous work~\cite{fang2023large,song2024code}, it has been reported to outperform other open-source models targeting code-related tasks (e.g., code generation).

    \item \textbf{Text-Embedding Models}~\cite{textemb2024} refer to a series of new-generation embedding models developed by OpenAI~\cite{openai2024}. 
    These models can generate both text and code embeddings with enhanced representation capabilities. 
    Specifically, we employed all three versions of text-embedding models (i.e., Text-Embedding-Ada-002, Text-Embedding-3-Small, and Text-Embedding-3-Large).
    
    \item \textbf{ChatGPT}~\cite{chatgpt2022} is a revolutionary LLM capable of transforming various fields, like software engineering.
    It is trained on large amounts of natural language text and code snippets, with reinforcement learning to follow human instructions.
    Particularly, we studied two LLMs (i.e., GPT-3.5-Turbo~\cite{chatgpt2022} and GPT-4~\cite{openai2023gpt4}). 
    % The potential of ChatGPT in code-related tasks has attracted significant attention in recent studies~\cite{liu2024your}.
\end{itemize}
To summarize, the studied LLMs can be divided into two types based on their architectures: encoder LLMs and decoder-only LLMs.
Encoder LLMs consist of encoder-only models (i.e., CodeBERT, GraphCodeBERT, Text-Embedding Models) and encoder-decoder models (i.e., PLBART, CodeT5, UniXCoder, CodeT5+, and StarCoder).
% Encoder LLMs comprise CodeBERT, GraphCodeBERT, LBART, CodeT5, UniXCoder, CodeT5+, StarCoder, and Text-Embedding Model (i.e., Text-Embedding-Ada-002, Text-Embedding-3-Small, and Text-Embedding-3-Large). 
Decoder-only LLMs include Code Llama and ChatGPT.
\subsection{Pre-trained Large Language Models for Code Embedding}
\label{subsec:approach}
Recently, pre-trained encoder LLMs have achieved substantial improvement in various code classification tasks, including code clone detection~\cite{khajezade2024investigating} and functionality classification~\cite{zhang2019novel}.
% , and vulnerability detection~\cite{ding2021towards}. 
Typically, these encoder LLMs are pre-trained on a large number of code snippets, learning the general-purpose code embedding knowledge. 
To adapt pre-trained LLMs to various downstream tasks, researchers usually train a multilayer perceptron (MLP) classifier to predict specific properties based on code embeddings produced by the pre-trained encoder LLMs~\cite{guo2020graphcodebert,niu2023empirical,tian2023fly}.
In our study, we denote this widely-used paradigm of LLMs as the \textbf{pre-trained code embedding strategy}, representing a fundamental LLM paradigm.

Hence, we also adopted the pre-trained code embedding strategy to our studied pre-trained encoder LLMs and investigated their effectiveness in detecting equivalent mutants. 
% based on the pre-training code embedding knowledge.
As shown in Figure~\ref{fig:overview}, we designed an encoder-based classifier framework, which consists of an LLM encoder and an MLP-based classifier.
Specifically, we first integrated the pre-trained encoder LLMs into our designed encoder-based classifier framework for training the domain-specific classifiers to detect equivalent mutants based on our training dataset.
During the training phase, we fixed the parameters of the pre-trained LLM encoder and only updated the parameters of the MLP-based classifier following existing work~\cite{guo2020graphcodebert,niu2023empirical}.
Subsequently, multiple training iterations are performed on the training data to enable the classifier to fully learn the detection of equivalent mutants utilizing the code embeddings generated by the pre-trained LLM encoder.

% \shl{First sentence "for encoder only LLMs" involving Text-Embedding-XXX, but in last paragraph of sec 3.2, we mentioned Text-Embedding-XXX series as embedding model. Looks like definition is conflict}
In particular, for encoder LLMs (i.e., CodeBERT, GraphCodeBERT, PLBART, CodeT5, UniXCoder, CodeT5+, StarCoder, Text-Embedding-Ada-002, Text-Embedding-3-Small, and Text-Embedding-3-Large), we utilized their encoder components for obtaining embedding vectors.
For the exceptional LLMs, Code Llama and ChatGPT, both of which are decoder-only architectures, are not applicable to this pre-trained code embedding strategy. 
% In Section~\ref{subsec:strategy}, we will further explore their performance in detecting equivalent mutants based on the other LLM strategies.
% More implementation details of these LLMs can be found in Section~\ref{subsec:implementation}. 
% In our experiment, all the open-source pre-trained models are downloaded from Huggingface~\cite{wolf2019huggingface}.
% While for three state-of-the-art embedding models provided by ChatGPT and ChatGPT, we used them through OpenAI’s APIs.
% We conducted all the experiments on an Intel Xeon CPU Gold-6342 machine with 512 GB RAM, Ubuntu 20.04.6, and two A800 GPUs.
% Similarly, we used the experimental implementation and environment for the experiments in Section~\ref{subsec:strategy}.
% More detail hyper-parameters can be found in our project homepage~\cite{EMD2024}. 

% \vspace{-2mm}
\subsection{Strategies for Large Language Models}
\label{subsec:strategy}
We further investigated the impact of various strategies using LLMs in equivalent mutant detection.
In Section~\ref{subsec:approach}, we initially presented the fundamental pre-trained code embedding strategy. 
Moreover, existing studies~\cite{liu2023pre,zhang2023revisiting} have demonstrated that fine-tuning strategies can effectively adapt general LLMs to specific downstream tasks, leading to significant enhancement in LLM performance.
Meanwhile, prompting strategies also have been proposed to achieve the same goal in a plug-and-play manner~\cite{kojima2022large}.
% Although they have achieved excellent performance in many downstream tasks, their effectiveness in equivalent mutant detection remains largely unexplored.
% Thus, evaluating the effects of various strategies can guide the selection of the most effective strategy and appropriate LLMs for detecting equivalent mutants.
Thus, we devised another four LLM strategies as follows:
\begin{itemize}[leftmargin=10pt]
    \item \textbf{Fine-tuned code embedding strategy}: we also employed the same encoder-based classifier framework and hyper-parameter settings as those applied in the \textbf{pre-trained code embedding strategy} across all the encoder LLMs.
    However, we did not fix the encoder parameters; instead, we simultaneously updated the parameters of both the encoder and classifier during training.
    
    \item \textbf{Zero-shot prompting strategy}: we devised a prompt without any examples, which directly utilized a mutant pair and a structured instruction (i.e., "Please identify if the two above codes are semantically equal. Please only answer `yes' or `no'. `yes' means they are semantically equal. `no' means they are not.") to prompt LLMs for equivalent mutant detection. 

    \item \textbf{Few-shot prompting strategy}: it enables LLMs to learn the relationship between the mutant pair and semantic equivalence based on randomly selected $<$\textit{mutant pair}, \textit{equivalence}$>$ examples. 
    That is, it concatenates these demonstration examples with a zero-shot prompt to form a new few-shot prompt, which is then fed to LLMs for equivalent mutant detection.
    
    \item \textbf{Fine-tuning with instruction strategy}: it enables LLMs to acquire specific knowledge through training on many more instruction-filled mutant pairs.
    Specifically, we used the same structured instruction that is described in the aforementioned zero-shot prompting strategy to construct an instruction-filled fine-tuning set, i.e., $<$\textit{mutant pair}, \textit{structured instruction}, \textit{equivalence}$>$. Then, we fine-tuned the LLMs on the fine-tuning set to detect equivalent mutants by the zero-shot prompt.
    % These instructions outline the task of equivalent mutant detection that we aim for the model to perform.
    % We applied same structured instruction in zero-shot prompt strategy and incorporated it with the mutant pairs to construct an instruction-filled training set (e.g., $<$\textit{instruction}, \textit{mutant pair}, \textit{equivalence}$>$).
    % Then, we fine-tuned LLMs on the training set to detect equivalent mutants by the prompt.

    % \wang{one sentence what is instruction}
    % To construct the fine-tuning set for LLMs, we combined the zero-shot prompt as the instruction for each mutant pair (e.g., $<$\textit{instruction}, \textit{mutant pair}, \textit{equivalence}$>$ in $MutantA$. 
    % Then, we fine-tuned the LLM on this build fine-tuning set to obtain a customized LLM. 
    % Finally, we used the customized LLM to predict the mutant pairs in $MutantB$.
    % \tz{it is still not clear. What is definition of the "instruction", "sample", "fine-tuning set", "customized LLM"? All these words seem to appear for the fist time}
    % \shl{it enables LLMs to acquire more knowledge about equivalent mutants by training on many more instruction filled samples, enhancing the LLM performance in equivalent mutant detection.
    % To construct the fine-tuning set for LLMs, we combine the zero-shot prompt as the instruction for each sample (e.g., $<$\textit{instruction}, \textit{mutant pair}, \textit{equivalence}$>$ in $MutantA$. 
    % Then, we fine-tune the LLM on this build fine-tuning set to obtain a customized LLM. 
    % Finally, we used the customized LLM to predict the mutant pairs in $MutantB$.
    
    % }
\end{itemize}

% \vspace{-2mm}
\subsection{Baselines}
\label{subsec:baselines}
To fairly evaluate LLM performance, we meticulously selected the baselines by conducting a succinct literature review of relevant papers published in SE venues over the last decade.
From this, we elected three widely studied techniques that rely solely on the code features without depending on the execution information of test cases, and are provided with a full replication package. 
These techniques encompass a total of ten baselines for comparison:

\begin{itemize}[leftmargin=10pt]
    \item \textbf{Compiler-based technique}.
    \textit{Trivial Compiler Equivalence (TCE)} is an EMD technique based on compilation optimization~\cite{papadakis2015trivial,kintis2017detecting}, employing the off-the-shelf compilers to compile the original program and each of its mutants into machine code, subsequently detecting mutant equivalence by comparing the equivalence of machine code pairs.
    % Due to its ease of implementation and superior performance, TCE has remained the most popular EMD technique for detecting equivalent mutants, widely used in previous work~\cite{chekam2019mart,ma2020commit,chekam2021killing,ma2021mudelta}.
    
    \item \textbf{ML-based technique}. Brito et al.~\cite{brito2020preliminary} extracted a set of features derived from source code properties and control flow information (e.g., mutation operator and graph distance).
    Based on these features, they then constructed seven ML classification models to detect equivalent mutants, including \textit{K-Nearest Neighbors (KNN)}, \textit{Decision Tree (DT)}, \textit{Random Forest (RF)}, \textit{Support Vector Machine (SVM)}, \textit{Linear Discriminant Analysis (LDA)}, \textit{Logistic Regression (LR)}, and \textit{Gaussian Naive Bayes (GNB)}.
    
    \item \textbf{Tree-based neural network technique}. 
    \textit{Abstract Syntax Tree Neural Network (ASTNN)} takes ASTs of mutant programs as input to detect mutant equivalence~\cite{peacock2021automatic}.
    This model can capture lexical-level and statement-level syntactical features of the code, as well as the code semantics by decomposing the large ASTs and encoding multi-way statement trees, significantly enhancing the detection performance.
    % and \textit{Tree-based Convolutional Neural Network (TBCNN)}~\cite{kusharki2022automatic}.
    % Specifically, \textit{ASTNN} model takes Abstract Syntax Trees ASTs as input to automatically detect equivalent mutants, while \textit{TBCNN} 
    % This model can capture both syntactic features of the code through the encoding of ASTs, as well as the semantics of the code by modeling the natural flow of statements, significantly enhancing the performance of detecting equivalent mutants.
    % Kusharki et al.~\cite{kusharki2022automatic} developed a model based on the \textit{Tree-based Convolutional Neural Network (TBCNN)} architecture to detect equivalent mutants.
    % This study demonstrated that, compared to ASTNN, this model more effectively utilized ASTs to capture syntactic and semantic features for equivalent mutant detection.
\end{itemize}

We replicated the baseline techniques by following the implementations and parameter settings recommended in previous papers.
% We followed the implementations and the parameter settings recommended by previous papers to replicate the baseline techniques.
Since the original version of the ML-based technique~\cite{brito2020preliminary} and the Tree-based NN technique~\cite{peacock2021automatic} only support C/C++ code, we expanded their capabilities to include Java code.
Moreover, we used two variants of the most widely-used TCE baseline, namely TCE$_{Javac}$ and TCE$_{Soot}$, for comparison.
% Specifically, TCE$_{Javac}$ runs the Javac compiler to compile the Java files of the mutant pair, resulting in corresponding binary class files for equivalent mutant detection. 
% In addition, TCE${Soot}$ runs the Soot framework~\cite{vallee2010soot} on each class file, which acts as the optimization phase of compilation.
Note that, we acknowledge the existence of code clone detection (CCD) techniques~\cite{khajezade2024investigating} that are built upon code similarities or patterns. 
However, EMD and CCD techniques serve distinct purposes and operate on different principles. 
Given that all mutants are inherently code snippets with minor syntactic changes (even an operator) from the original program, all the mutant pairs can be considered code clones.
Therefore, we opted not to include CCD techniques as baselines in our study.
% following existing work~\cite{papadakis2015trivial,kintis2017detecting,brito2020preliminary,peacock2021automatic}. 
% In the future, we aim to further investigate the performance EMD techniques on code clone detection, and vice versa.}

% \vspace{-2mm}
\subsection{Metrics}
\label{subsec:metrics}

\

\textbf{Effectiveness}.
Following existing work~\cite{naeem2020machine,brito2020preliminary,chung2022augmenting,peacock2021automatic}, we adopted the most widely-used metrics for the binary classification task, \textit{Precision}, \textit{Recall}, and \textit{F1-score}, to assess the effectiveness of all our studied EMD techniques for equivalent mutant detection.
The F1-score, being the harmonic mean of recall and precision, offers a balanced assessment of detection performance.
% Specifically, \textit{Precision} is computed by $\frac{TP}{TP+FP}$ while \textit{Recall} is computed by $\frac{TP}{TP+FN}$, where $TP$, $FP$, and $FN$ refer to the number of true positives (a mutant is predicted to be equivalent and its ground truth is also equivalent), false positives (a mutant is predicted to be equivalent but its ground truth is not equivalent), and false negatives (a mutant is predicted to be not equivalent but its ground truth is equivalent), respectively. 
% % \textit{F1-Score} considers both \textit{Precision} and \textit{Recall}, which is calculateed by $\frac{2*(Precision*Recall)}{Precision+Recall}$.
% The F1-Score, being the harmonic mean of recall and precision, offers a balanced assessment of model performance by considering both metrics, calculated by $\frac{2*(Precision*Recall)}{Precision+Recall}$. 
% \shl{we used Macro average instead of using Binary average, to update this part}
Specifically, in our study, we used macro-averaged precision, recall, and F1-score metrics.
These macro-averaged metrics are unbiased by potential class imbalances, which are calculated by finding the unweighted mean of the respective metrics for each class. 
For example, given two labels (i.e., negative and positive), the macro-average F1-score is the average of the F1-score for both classes (i.e., $\frac{F1_{positive}+F1_{negative}}{2}$).

% Therefore, we have chosen the F1-Score as our primary evaluation metric for assessing the effectiveness of each model.

% The recall of a model measures its ability to correctly identify instances of the positive class (i.e., equivalent mutants in this study), with a higher recall indicating fewer false negatives.
% Precision reflects the proportion of positive predictions made by the model that are correct, with a high precision indicating a low false positive rate. 
% Conversely, a low precision suggests that the model is incorrectly predicting many items as positive. 
% The F1-Score, being the harmonic mean of recall and precision, offers a balanced assessment of model performance by considering both metrics. 
% Therefore, we have chosen the F1-Score as our primary evaluation metric for assessing the effectiveness of each model.

\textbf{Efficiency}. In practice, the time spent on detecting a mutant pair is significant due to the large number of generated mutants in the mutation testing scenario.
We compared the time overheads among the techniques on equivalent mutant detection to quantitatively measure their efficiency. 
Two types of time overheads are defined: the average time spent on 
detecting a mutant pair (referred to as \textit{inference time}) and the total time spent building an EMD model based on the training set offline (referred to as \textit{training time}).
% We call the former \textit{average inference time} and the latter \textit{total training time}.

% \vspace{-2mm}
\subsection{Implementation and Environment}
\label{subsec:implementation}
% We implement the generation pipeline in Python using specific PyTorch versions of each LLM.
All the open-source pre-trained models (i.e., CodeBERT, GraphCodeBERT, PLBART, CodeT5, UniXCoder, CodeT5+, and StarCoder) are downloaded from Huggingface~\cite{wolf2019huggingface}.
In particular, we used Text-Embedding-Ada-002, Text-Embedding-3-Small, Text-Embedding-3-Large, GPT-3.5-Turbo, and GPT-4 through OpenAI’s APIs~\cite{openai2024}.
Concretely, we used \texttt{gpt-3.5-turbo-0125} as the specific experimental model version for GPT-3.5-Turbo and used \texttt{gpt-4-0613} as the experimental model version for GPT-4.
We utilized the recommended hyper-parameters~\cite{lu2021codexglue} for the pre-trained code embedding strategy due to their proven effectiveness.
To ensure a fair comparison and reduce the complexity of hyper-parameter tuning, we reused the same hyper-parameters for the fine-tuned code embedding strategy. 
More detailed settings of hyper-parameters can be found in our project homepage~\cite{EMD2024}.
% \tz{The specific versions of GPT-3.5 and GPT-4 is ? for example gpt-3.5-0301}
% To promote practical use and future research, we have released both the implementation and the hyper-parameter settings for each LLM model at our project homepage~\cite{?}. 
% All the hyper-parameter settings for each models could be found in our anonymous homepage~\cite{?}.
% Regarding baselines, we directly employed their open-source implementations alongside the
% recommended configurations for comparison.
We conducted all the experiments on an Intel Xeon CPU Gold-6342 machine with 512 GB RAM, Ubuntu 20.04.6, and two A800 GPUs.

\section{Results}
\label{sec:results_and_analysis}

\subsection{RQ1: Performance of LLMs in EMD}
\label{subsec:RQ1}
\noindent
\textbf{\emph{\underline{Approach.}}}
This research question offers a comparative analysis of the performance of various encoder LLMs, specifically focusing on their usage of code embeddings.
The typical pre-trained code embedding strategy of LLMs for equivalent mutant detection has been provided in Section~\ref{subsec:approach}.
For all eight ML-based and Tree-based NN baselines, we also trained their classifiers with our training dataset based on the same settings and process as their corresponding paper.
The exceptional Compiler-based baselines (i.e., TCE$_{Javac}$ and TCE$_{Soot}$) are based on off-the-shelf compilers, requiring no training phase.
Subsequently, we measured the effectiveness of 10 state-of-the-art encoder LLMs and 10 baselines in terms of precision, recall, and F1-score metrics.
% To assess the statistical significance of these correlations, we conducted a t-test to compute p-values.

\noindent
\textbf{\emph{\underline{Results.}}}
Table~\ref{tab:rq1} shows the comparison results among LLMs and the baselines in terms of precision, recall, and F1-score.
First, the evaluation results show that almost all LLM-based techniques (except CodeBERT) achieve superior effectiveness compared to all the baselines in terms of F1-score.
For example, the most effective LLM-based techniques (i.e., UniXCoder and CodeT5+) achieve the F1-score of 81.88\%, whereas the F1-score of ASTNN, KNN, and TCE$_{Soot}$ is 70.00\%, 72.15\%, and 50.80\%, respectively. 
On average, LLM-based techniques outperform the Compiler-based, ML-based, and Tree-based NN techniques by 75.18\%, 19.14\%, and 12.75\% in terms of F1-score, 108.27\%, 15.90\%, and 8.23\% in terms of precision, and 62.05\%, 15.25\%, and 11.68\% in terms of recall, respectively.
% Furthermore, the statistical tests have confirmed a significant difference between the studied LLMs and all the baselines. \wang{where describe this statistical testing?}
It significantly demonstrates the effectiveness of LLMs on equivalent mutant detection.
The prevalence of this phenomenon may stem from the fact that LLMs, typically pre-trained on extensive code snippets, exhibit enhanced capability in handling code-related downstream tasks compared to the general ML/DL models lacking such pre-training strategies.
In particular, through manual analysis, we found that 363 mutant pairs could not be compiled successfully by Javac due to missing necessary configuration files. 
Since both TCE$_{Javac}$ and TCE$_{Soot}$ rely on the corresponding binary classfiles produced by Javac, their performance is consequently suboptimal.
% Both TCE$_{Javac}$ and TCE$_{Soot}$ leverage Javac compiler to compiles the Java files of the mutant pair into the corresponding binary classfiles before detection.

Second, we observe that the performance of three state-of-the-art Text-Embedding models does not exceed other pre-trained encoder LLMs.
Conversely, UniXCoder and CodeT5+, characterized by fewer parameters, demonstrate relatively superior performance in equivalent mutant detection. 
For instance, the F1-score of Text-Embedding-Ada-002 is 74.56\% while that of UniXCoder is 81.88\%.
This may primarily arise because embedding models are general-purpose text embedding models trained on extensive natural language datasets. 
However, when employed for the code-related task (i.e., equivalent mutant detection), they are indeed affected by the data-shift problem~\cite{ma2023scope}.
It also suggests that smaller code-specific encoder LLMs are more inclined to produce code representations conducive to MLP-based classifiers learning the semantic differences in code.
% This process aids in learning vital semantic information necessary for equivalent mutant detection.

% Third, among these baselines, KNN outperforms traditional Compiler-based techniques (i.e., TCE$_{Javac}$ and TCE$_{Soot}$), other ML-based techniques (i.e., DT, RF, SVM, LDA, LR, and GNB), and even more advanced Tree-based NN technique (i.e., ASTNN) in terms of F1-score.
% More specifically, KNN achieves an improvement of 2.15\%$\sim$83.54\% in the F1-score across all other baselines.
% This finding indicates that the superior effectiveness of ML models in detecting equivalent mutants based on appropriate designed mutant features.

% Fourth, when comparing TCE$_{Javac}$ to TCE$_{Soot}$, the latter demonstrates significantly superior performance. 
% This is because TCE$_{Javac}$ just runs Javac compiler to compiles the Java files of the mutant pair, which lead to corresponding classfiles.
% Due to the limited optimization rules, TCE$_{Javac}$ cannot detect complex equivalent mutants that the mutation operators are not belong to optimization rules. 
% While TCE$_{Soot}$ then additionally runs Soot framework~\cite{vallee2010soot} on each classfile, acting as the optimization phase of compilation, in order to discover equivalences between programs that normal Javac compiler would not detect.

% \vspace{-2mm}
\begin{tcolorbox}\textbf{RQ1 Summary:}
LLMs significantly surpass all ten EMD baselines in equivalent mutant detection.
Specifically, LLMs outperform the Compiler-based, ML-based, and Tree-based NN techniques by an average improvement of 75.18\%, 19.14\%, and 12.75\% in terms of F1-score, respectively.
\end{tcolorbox}

\begin{table}[t]
    \caption{The performance of baselines and state-of-the-art LLMs on equivalent mutant detection}
    \label{tab:rq1}
    \centering
    \tabcolsep=3.0mm
    % \vspace{-.2cm}
    \small
    \begin{adjustbox}{max width=1.0\textwidth, center}
        \begin{tabular}{lrrr}
            \toprule
            \toprule
            \textbf{Technique} & \textbf{Precision} & \textbf{Recall} & \textbf{F1-score} \\ 
        	\midrule
            \rowcolor{mygray}\multicolumn{4}{l}{\textbf{Compiler-based technique}} \\ \midrule
            TCE$_{Javac}$ & 40.55\% & 38.15\% & 39.31\% \\
            TCE$_{Soot}$ & 51.26\% & 51.81\% & 50.80\% \\ \midrule
            \rowcolor{mygray}\multicolumn{4}{l}{\textbf{ML-based technique}} \\ \midrule
            KNN & 77.77\% & 69.09\% & 72.15\% \\
            DT & 78.20\% & 67.20\% & 70.63\% \\
            RF & 78.89\% & 66.94\% & 70.53\% \\
            SVM & 93.62\% & 58.84\% & 61.61\% \\
            LDA & 88.02\% & 59.22\% & 62.09\% \\
            LR & 90.69\% & 59.33\% & 62.30\% \\
            GNB & 70.23\% & 62.10\% & 64.43\% \\ \midrule
            \rowcolor{mygray}\multicolumn{4}{l}{\textbf{Tree-based NN technique}} \\ \midrule
            ASTNN & 88.34\% & 65.27\% & 70.00\% \\ \midrule
            % Kusharki et al.\cite{kusharki2022automatic} & TBCNN & 90.05\% & 69.29\% & 74.62\% \\ \midrule
            \rowcolor{mygray}\multicolumn{4}{l}{\textbf{LLM-based technique (Pre-trained code embedding strategy)}} \\ \midrule
            CodeBERT (110M) & 94.36\% & 64.26\% & 69.20\% \\
            GraphCodeBERT (110M) & 95.81\% & 74.30\% & 80.52\% \\
            PLBART (210M) & 95.81\% & 74.30\% & 80.52\% \\
            CodeT5 (210M) & 95.81\% & 74.30\% & 80.52\% \\
            UniXCoder (110M) & \textbf{96.02\%} & \textbf{75.70\%} & \textbf{81.88\%} \\
            CodeT5+ (6B) & \textbf{96.02\%} & \textbf{75.70\%} & \textbf{81.88\%} \\
            StarCoder (7B) & 95.99\% & 75.50\% & 81.69\% \\
            % \multicolumn{2}{l}{Code Llama} & 100.00\% & 51.41\% & 67.90\% \\
            Text-Embedding-Ada-002 & 94.99\% & 68.67\% & 74.56\% \\
            Text-Embedding-3-Small & 95.31\% & 70.88\% & 77.00\% \\
            Text-Embedding-3-Large & 95.96\% & 75.30\% & 81.50\% \\
            \bottomrule
            \bottomrule
        \end{tabular}
    \end{adjustbox}
    % \vspace{-.5cm}
\end{table}

\subsection{RQ2: Best Strategy of LLMs in EMD}
\label{subsec:RQ2}
\noindent
\textbf{\emph{\underline{Approach.}}}
This research question aims to investigate the impact of four additional LLM strategies (designed in Section~\ref{subsec:strategy}) for enhancing equivalent mutant detection compared to the pre-trained code embedding strategy in RQ1.
% \wang{please check}
Similarly, for the fine-tuned code embedding strategy, we selected CodeBERT, GraphCodeBERT, PLBART, CodeT5, UniXCoder, CodeT5+, and StarCoder as the base models. 
%\shl{Note removed Text-Embedding here}
Besides, we selected Code Llama, GPT-3.5-Turbo, and GPT-4 as the representative LLMs to perform zero-shot prompting, few-shot prompting, and fine-tuning with instruction strategies.
Given the input length limit of LLMs, we used the 3-shot setting for the few-shot prompting strategy following the existing work~\cite{ma2023scope}.
In particular, we excluded GPT-4 from the fine-tuning with instruction strategy due to its unavailability.
Specifically, we examined the extent of impact resulting from these four LLM strategies in terms of precision, recall, and F1-score metrics.
% To assess the statistical significance of these correlations, we also conducted a t-test to compute p-values.

\noindent
\textbf{\emph{\underline{Results.}}}
Table~\ref{tab:rq1} and Table~\ref{tab:rq2} illustrate the performance of five LLM strategies on equivalent mutant detection.
First, the fine-tuned UniXCoder achieves the best performance compared to all other combinations of LLMs and strategies with the improvement of 1.16\%$\sim$78.85\% in terms of F1-score.
It demonstrates employing smaller LLMs with the fine-tuned code embedding strategy is the best approach for equivalent mutant detection.

Second, the fine-tuned code embedding strategy always outperforms the pre-trained code embedding strategy on all studied LLMs in terms of F1-score.
For example, the former outperforms the latter with the improvement of 21.20\%, 5.79\%, 6.09\%, 4.78\%, 5.74\%, 4.53\%, and 0.23\% on CodeBERT, GraphCodeBERT, PLBART, CodeT5, UniXCoder, CodeT5+, and StarCoder in terms of F1-score, respectively.
It indicates that the fine-tuned code embedding strategy can significantly improve the LLM performance in equivalent mutant detection. 
% Besides, our manual analysis revealed that despite the superiority of fine-tuned LLMs over pre-trained LLMs in overall performance, a fraction of pre-training knowledge is still lost during the fine-tuning process. 
Additionally, upon inspecting their prediction results, we observed that despite the superiority of fine-tuned LLMs over pre-trained LLMs in overall performance, a fraction of pre-training knowledge is still lost during the fine-tuning process.
Specifically, among the initially correct predictions, 1.09\% were erroneously transformed into incorrect predictions on average.
This observation exposes the catastrophic forgetting problem~\cite{huang2023empirical} of LLMs under the fine-tuning paradigm to some extent.

Third, code embedding strategies (i.e., pre-trained and fine-tuned code embedding) outperform prompting strategies (i.e., zero-shot and few-shot prompting) across all three metrics (i.e., precision, recall, and F1-score).
On average, code embedding strategies outperform prompting strategies by 55.81\%, 41.50\%, and 54.21\% in terms of precision, recall, and F1-score, respectively.
The main reason lies in the inherent complexity of context-understanding demanded by LLMs (with prompting strategies) for the comprehensive understanding of mutant pairs, compared to the relatively simplified process of directly comparing the embedding vectors of mutant pairs. 
As a result, code embedding strategies appear more straightforward in mutant understanding and comparison, thereby leading to superior detection performance.

% Third, while few-shot prompting can enhance the performance of GPT-4, it may result in decreased performance for Code Llama and GPT-3.5-Turbo when compared to zero-shot prompting.
% For instance, GPT-4 + few-shot prompting outperforms GPT-4 + zero-shot prompting by 4.27\% in terms of F1-score. Conversely, zero-shot prompting exhibits superior performance over few-shot prompting for Code Llama and GPT-3.5-Turbo, with an average improvement of xxxx\%\tz{todo} in terms of F1-score.
% These findings suggest that demonstration examples provided by few-shot prompting do not consistently enhance the performance of LLMs for detecting equivalent mutants.
% ~\cite{wang2023towards,ma2023scope}

Fourth, fine-tuning with instruction strategy significantly outperforms both zero-shot prompting and few-shot prompting strategies based on both decoder-only LLMs (i.e., Code Llama and GPT-3.5-Turbo) across all three metrics.
On average, fine-tuning with instruction strategy improves 57.07\% and 69.55\% higher precision than zero-shot prompting and few-shot prompting, 19.41\% and 29.06\% higher recall, and 28.34\% and 39.23\% higher F1-score, respectively.
It further demonstrates that the fine-tuning strategy can significantly enhance the performance of LLMs in equivalent mutant detection.

\begin{table}[t]
    \caption{The performance of different LLM strategies on equivalent mutant detection}
    % \vspace{-2mm}
    \label{tab:rq2}
    \centering
    \tabcolsep=3.2mm
    \small
    \begin{adjustbox}{max width=1.0 \textwidth,center}
        \begin{tabular}{ lrrr }
            \toprule
            \toprule
            \textbf{Technique} & \textbf{Precision} & \textbf{Recall} & \textbf{F1-score}  \\ \midrule
            \rowcolor{mygray}\multicolumn{4}{l}{\textbf{Fine-tuned code embedding strategy}} \\ \midrule
            CodeBERT (110M) & 90.39\% & 79.74\% & 83.87\% \\
            GraphCodeBERT (110M) & 91.54\% & 81.05\% & 85.18\% \\
            PLBART (210M) & 93.24\% & 80.70\% & 85.42\% \\
            CodeT5 (210M) & 90.59\% & 80.34\% & 84.37\% \\
            UniXCoder (110M) & 94.33\% & 81.81\% & \textbf{86.58\%} \\
            CodeT5+ (6B) & 89.28\% & \textbf{82.79\%} & 85.59\% \\
            StarCoder (7B) & \textbf{96.02\%} & 75.70\% & 81.88\% \\ \midrule
            \rowcolor{mygray}\multicolumn{4}{l}{\textbf{Zero-shot prompting strategy}} \\ \midrule
            Code Llama (7B) & 59.22\% & 50.78\% & 48.04\% \\ 
            GPT-3.5-Turbo & 59.22\% & 59.70\% & 59.44\% \\
            GPT-4 & 67.42\% & 53.76\% & 53.61\% \\ \midrule
            \rowcolor{mygray}\multicolumn{4}{l}{\textbf{Few-shot prompting strategy}} \\ \midrule
            Code Llama (7B)  & 52.85\% & 50.38\% & 47.76\% \\ 
            GPT-3.5-Turbo & 57.04\% & 52.23\% & 51.59\% \\ 
            GPT-4 & 67.02\% & 55.18\% & 55.90\% \\ \midrule
            \rowcolor{mygray}\multicolumn{4}{l}{\textbf{Fine-tuning with instruction strategy}} \\ \midrule
            Code Llama (7B) & 93.21\% & 55.82\% & 56.79\% \\
            GPT-3.5-Turbo & 92.82\% & 76.95\% & 82.31\% \\
            \bottomrule
            \bottomrule
        \end{tabular}
    \end{adjustbox}
    % \vspace{-.5cm}
\end{table}

\begin{tcolorbox}\textbf{RQ2 Summary:}
The fine-tuned UniXCoder significantly outperforms all other combinations of LLMs and strategies with the improvement of 1.16\%$\sim$78.85\% in terms of F1-score, demonstrating that fine-tuned code embedding strategy is the best strategy on equivalent mutant detection.
Additionally, LLMs based solely on prompting strategies cannot achieve comparable performance.
\end{tcolorbox}

\begin{figure*}[t!]
    \centering
    \includegraphics[width=1.0\linewidth]{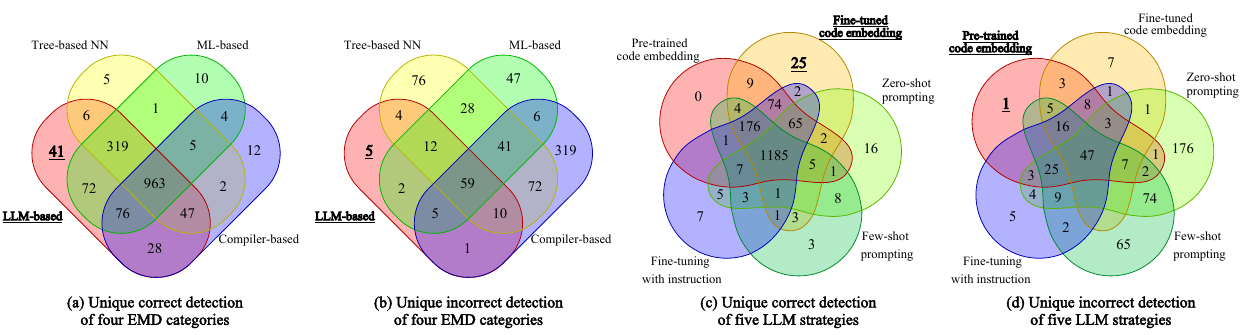}
    % \vspace{-4mm}
    \caption{Unique correct detections ($\uparrow$) and unique incorrect detections ($\downarrow$) across studied EMD techniques}
    \label{fig:RQ3_veen}
    % \vspace{-.5cm}
    % \vspace{-4mm}
\end{figure*}

% \vspace{-3mm}
\subsection{RQ3: Orthogonality between Studied EMD Techniques}
\label{subsec:RQ3}
\noindent
\textbf{\emph{\underline{Approach.}}} 
This research question aims to gain an understanding of the performance characteristics of different EMD techniques and LLM strategies.
Hence, we conducted a comprehensive analysis to explore the degree of their orthogonality with two different perspectives, following the experimental design of RQ1 and RQ2:
\begin{itemize}[leftmargin=10pt]
    \item \textbf{Between EMD categories.} Based on the findings of RQ1, we selected the best-performing EMD techniques within four EMD categories (i.e., Compiler-based, ML-based, Tree-based NN, and LLM-based techniques).
    Specifically, the selected techniques are TCE$_{Soot}$, KNN, ASTNN, and fine-tuned UniXCoder, each representing their respective EMD categories.
    
    \item \textbf{Between LLM strategies.} Based on the findings of RQ2, we selected the best-performing EMD techniques within five LLM strategies (i.e., pre-trained code embedding, fine-tuned code embedding, zero-shot prompting, few-shot prompting, and fine-tuning with instruction strategies).
    Specifically, the selected techniques are pre-trained UniXCoder, fine-tuned UniXCoder, ``GPT-3.5-Turbo + zero-shot prompting'', ``GPT-4 + few-shot prompting'', and ``GPT-3.5-Turbo + fine-tuning with instruction'', each representing their respective LLM strategies. 
\end{itemize}
% \vspace{-.2cm}

% \vspace{-1mm}
Based on the above two perspectives, we further conducted a two-level analysis: (1) \textbf{the unique correct/incorrect detections}, and (2) \textbf{the detection performance on each mutation operator}.
Regarding the first level, we employed Venn diagrams to assess the unique correct/incorrect detections across various studied EMD techniques. 
Regarding the second level, we further investigated the detection performance of each EMD technique across various mutation operators,
by disaggregating the detection results based on their respective mutation operators.
In particular, the test dataset comprises 1,611 mutants, each labeled with mutation operators, alongside an additional 39 mutants unlabeled.
Then the first two authors conducted a round-table discussion to manually classify these unlabeled mutants, aligning with the definition of mutation operators provided by the prior work~\cite{van2021mutantbench}.
Furthermore, we conducted a \textit{Kruskal-Wallis test}~\cite{Kruskal-wallis-test}, a non-parametric test for comparing differences among multiple independent groups, to assess the statistical significance between our studied EMD techniques in terms of the detection performance on each mutation operator.

\begin{figure}[]
    \centering
    \subfloat[Performance of 4 EMD categories on Top-10 mutation operators]{
        \includegraphics[width=1.\linewidth]{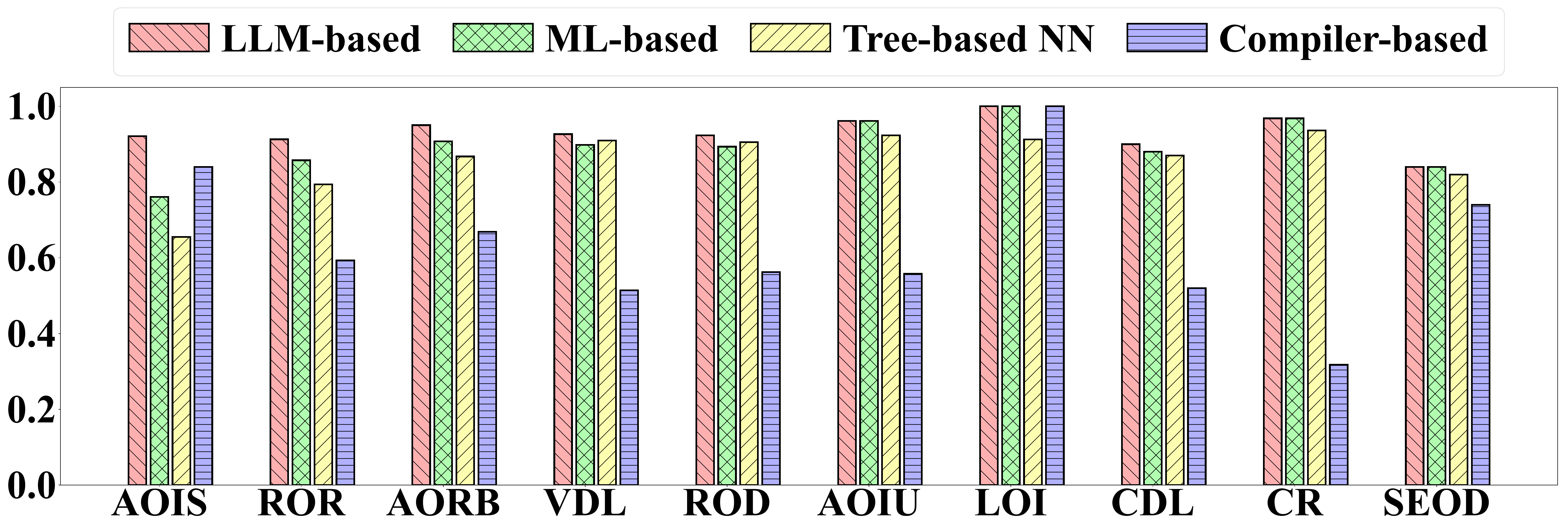}
        \label{fig:RQ3barplot1}
    }\\
    % \vspace{-2mm}
    \subfloat[Performance of 5 LLM strategies on Top-10 mutation operators]{
        \includegraphics[width=1.\linewidth]{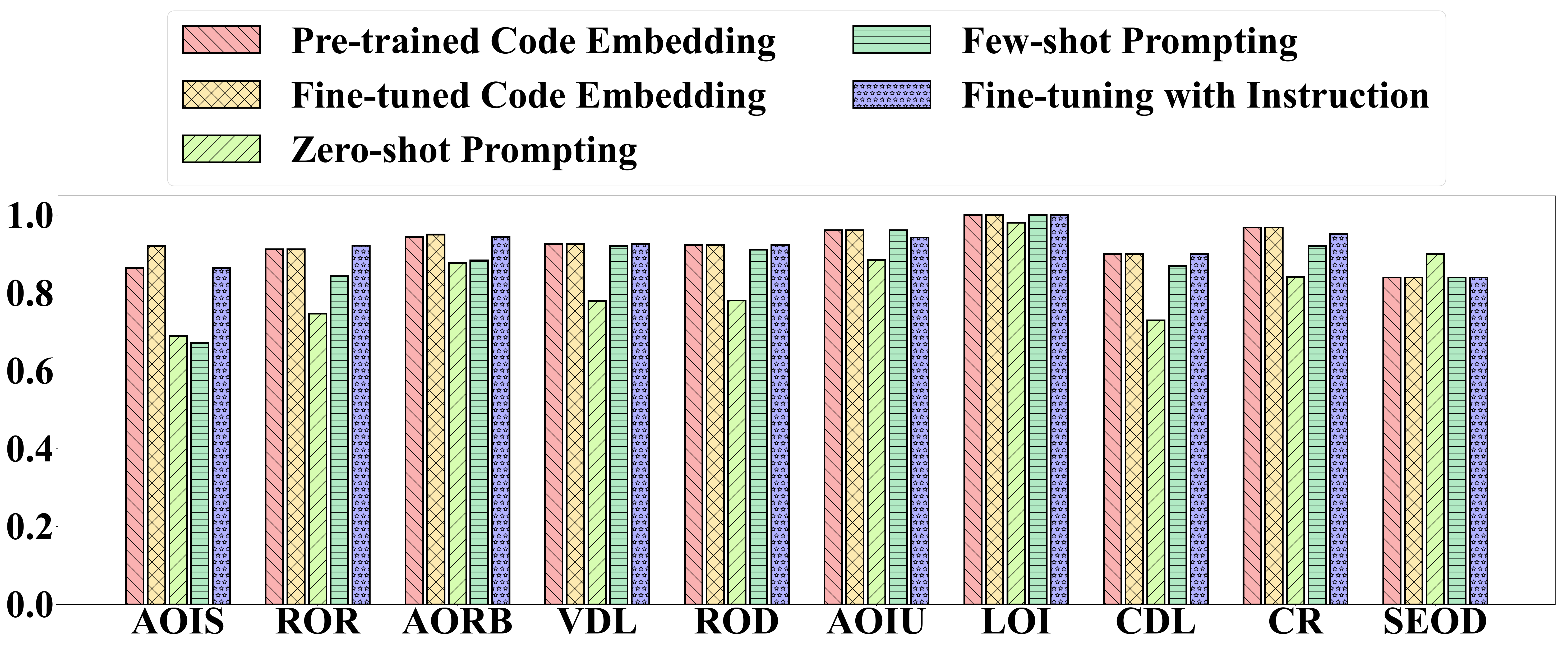}
        \label{fig:RQ3barplot2}
    }
    % \vspace{-2mm}
    \caption{Detection performance on Top-10 mutation operators across various EMD techniques (x-axis shows mutation operators and y-axis shows the correct detection percentage)}
    \label{fig:RQ3barplot}
    % \vspace{-5mm}
\end{figure}

\noindent
\textbf{\emph{\underline{Results.}}}
Figure~\ref{fig:RQ3_veen} presents the Venn diagrams that demonstrate the intersection of correct/incorrect detections among the studied EMD techniques based on two analyzed perspectives (i.e., EMD categories and LLM strategies).
Overlap areas denote shared correct/incorrect detections among multiple EMD techniques, while non-overlapping areas signify the unique correct/incorrect detections of each EMD technique.
Figure~\ref{fig:RQ3barplot} further shows the detection performance of studied EMD techniques on each mutation operator.
Due to space constraints, we only present the results on the top 10 common mutation operators. 
Detailed results for all 28 mutation operators are available on our project homepage~\cite{EMD2024}.

% \begin{figure}[]
%     \centering
%     \subfloat[Performance of 4 EMD categories on Top-10 mutation operators]{
%         % \includegraphics[width=1.\linewidth]{figures/invocation.pdf}
%         \includegraphics[width=1.\linewidth]{figures/trail-1.pdf}
%         % \includegraphics[width=1.\linewidth]{figures/4ways_bar_plot.pdf}
%         \label{fig:RQ3barplot1}
%     }\\
%     % \vspace{-2mm}
%     \subfloat[Performance of 5 LLM strategies on Top-10 mutation operators]{
%         \includegraphics[width=1.\linewidth]{figures/trail-2.pdf}
%         % \includegraphics[width=1.\linewidth]{figures/5ways_bar_plot.pdf}
%         \label{fig:RQ3barplot2}
%     }
%     % \vspace{-2mm}
%     \caption{Detection performance on Top-10 mutation operators across various EMD techniques (x-axis shows mutation operators and y-axis shows the correct detection percentage)}
%     \label{fig:RQ3barplot}
%     % \vspace{-5mm}
% \end{figure}

\textbf{Between EMD categories.}
From Figure~\ref{fig:RQ3_veen}a, we find that the LLM-based technique achieves the best performance compared to the other three EMD categories in terms of the unique \textit{correct} detections. 
Specifically, the LLM-based technique exhibits 41 unique correct detections, significantly surpassing 12 by the Compiler-based technique, 10 by the ML-based technique, and 5 by the Tree-based NN technique, respectively.
From Figure~\ref{fig:RQ3_veen}b, we find that the LLM-based technique also outperforms the other three EMD categories in terms of unique \textit{incorrect} detections. 
It only has 5 unique incorrect detections, which is significantly fewer than the other EMD categories (319, 47, and 76, separately).
These results prove the effectiveness of LLM-based techniques in equivalent mutant detection, further strengthening the findings in RQ1.

From Figure~\ref{fig:RQ3barplot1}, we further find that the LLM-based technique always outperforms the other three EMD categories in terms of detection performance on almost all mutation operators (with only one exception).
For example, the LLM-based technique achieves correct detections of 339 (92.12\%), 314 (91.28\%), and 287 (95.03\%) on the three most common mutation operators (i.e., \texttt{AOIS}, \texttt{ROR}, and \texttt{AROB}), while the suboptimal ML-based technique achieves lower correct detections of 280 (76.09\%), 295 (85.76\%), and 274 (90.73\%), respectively.
For the one exceptional mutation operator (i.e., \texttt{COR}), the LLM-based technique exhibits slightly fewer correct detections compared to the most effective ML-based technique (13 vs 14).
% Besides, we find that mutant pairs with ABSI operator pose a challenge for Tree-based NN technique (2/9), ML-based technique (4/9), and Compiler-based technique (0/9), whereas LLM-based technique can detect all of them. \tz{since the number of mutants with ABSI is too small, the conclusion is not convincing enough.}
Furthermore, the \textit{Kruskal-Wallis test} confirms a significant difference, with a \textit{p-value} of 8.89e-4, suggesting that the LLM-based technique is statistically superior to all the compared EMD categories in terms of the detection performance on each mutation operator.

% Furthermore, the Kruskal-Wallis test confirms that there exists a significant difference between EMD technique categories, with a \textit{p-value} $<$ 0.001.
% These results suggest that the LLM-based technique can effectively cover most mutant pairs regardless of their operator types and result in fewer incorrect detections.
% We obtained a p-value of 0.008893, which confirm that the difference in performance were statistically significant.

\textbf{Between LLM strategies.}
From Figure~\ref{fig:RQ3_veen}c, we find that the fine-tuned code embedding strategy performs best compared to the other four LLM strategies in terms of the unique \textit{correct} detections. 
Specifically, the fine-tuned code embedding strategy exhibits 25 unique correct detections, significantly surpassing 0 by the pre-trained code embedding strategy, 16 by the zero-shot prompting strategy, 3 by the few-shot prompting strategy, and 7 by the fine-tuning with instruction strategy.
From Figure~\ref{fig:RQ3_veen}d, we find that both prompting strategies achieve the poorest performance among five LLM strategies regarding the unique \textit{incorrect} detections, with 176/65 by the zero-shot/few-shot prompting strategy, compared to 1 by the pre-trained code embedding strategy, 7 by the fine-tuned code embedding strategy, and 5 by the fine-tuning with instruction strategy.
It significantly suggests that LLMs based solely on prompting strategies cannot achieve comparable performance on equivalent mutant detection, further supporting the findings in RQ2.
% Besides, pre-trained code embedding strategy outperforms fine-tuned code embedding strategy in terms of the unique incorrect detections, thus further confirming the catastrophic forgetting problem of LLMs as identified in RQ2.
%wang: I don't want to make it repetitive. we can later discuss it more in the discussion. 

From Figure~\ref{fig:RQ3barplot2}, we observe that the fine-tuned code embedding strategy always outperforms the other strategies in terms of detection performance on almost all mutation operators (with only two exceptions).
%(except in two cases). 
For the two exceptional mutation operators (i.e., \texttt{SEOD} and \texttt{ROR}), the fine-tuned code embedding strategy exhibits slightly fewer correct detections compared to the most effective strategies (42 vs 45 and 314 vs 317, respectively).
% zero-shot prompting and fine-tuning with instruction 
Besides, we find that both zero-shot and few-shot prompting strategies exhibit the lowest detection performance on most mutation operators.
For example, the zero-shot prompting strategy achieves correct detections of 254 (69.02\%), 257 (74.71\%), and 265 (87.75\%) on the three most common mutation operators (i.e., \texttt{AOIS}, \texttt{ROR}, and \texttt{AROB}), while the fine-tuned code embedding strategy achieves higher correct detections of 339 (92.12\%), 314 (91.28\%), and 287 (95.03\%) respectively.
Additionally, the \textit{Kruskal-Wallis test} validates that there is a significant difference among all the compared strategies in terms of detection performance on each mutation operator, with a \textit{p-value} being 1.63e-4, suggesting the superiority of the fine-tuned code embedding strategy.

% \vspace{-2mm}
\begin{tcolorbox}\textbf{RQ3 Summary:}
The LLM-based technique and the fine-tuned code embedding strategy significantly surpass the other EMD categories and LLM strategies regarding the unique correct/incorrect detections and the detection performance on each mutation operator.
\end{tcolorbox}

\subsection{RQ4: Efficiency of Studied EMD Techniques}
\label{subsec:RQ4}
\noindent
\textbf{\emph{\underline{Approach.}}}
% With the recent exponential growth in the size of LLMs, the associated computational costs have risen substantially. 
% Although larger LLMs may exhibit superior performance, it is imperative to consider the trade-off between their detection capabilities and the incurred costs.
This research question aims to assess the efficiency of our studied EMD techniques by calculating both training time (the total time spent building an EMD model offline) and inference time (the average time spent detecting a mutant pair).
Given that the training phase is not universally applicable to all EMD techniques, such as TCE, and is conducted offline only once before the inference phase, the primary metric for evaluating efficiency across these techniques is the inference time.

\begin{table}[t]
    \caption{Time efficiency of studied EMD techniques}
    % \vspace{-2mm}
    \label{tab:rq4}
    \centering
    \tabcolsep=1.3mm
    \small
    \begin{adjustbox}{max width=1.0 \textwidth,center}
        \begin{tabular}{ llrr }
            \toprule
            \toprule
            \multicolumn{2}{c}{\textbf{Technique}} & \textbf{\makecell{Training \\ Time (s)}} & \textbf{\makecell{Inference \\ Time (s)}} \\ 
        	\midrule
            \multirow{2}{*}{\textbf{Compiler-based}} & TCE$_{Javac}$ & - & 1.0241 \\
            & TCE$_{Soot}$ & - & 2.3537 \\ \midrule
            \multirow{7}{*}{\textbf{ML-based}} & KNN & 298.8415 & 0.0019 \\
            & DT & 297.3026 & 0.0015 \\
            & RF & 300.3978 & 0.0081 \\
            & SVM & 297.4997 & 0.0018 \\
            & LDA & 297.7096 & 0.0016 \\
            & LR & 296.8087 & 0.0016 \\
            & GNB & 298.2195 & 0.0014 \\ \midrule
            \textbf{Tree-based NN} & ASTNN & 306.7047 & 0.0274 \\ \midrule
            \rowcolor{mygray}\multicolumn{4}{c}{\textbf{LLM-based Technique}} \\ \midrule
            \multirow{10}{*}{\textbf{\makecell[l]{Pre-trained \\ code embedding}}} 
            & CodeBERT (110M) & 562.6160 & 0.0269 \\
            & GraphCodeBERT (110M) & 805.1435 & 0.0429 \\
            & PLBART (210M) & 844.1389 & 0.0421 \\
            & CodeT5 (210M) & 1545.3771 & 0.0784 \\
            & UniXCoder (110M) & 809.1785 & 0.0431 \\
            & CodeT5+ (6B) & 17043.0572 & 0.8294 \\
            & StarCoder (7B) & 16634.3038 & 0.9292 \\ 
            & Text-Embedding-Ada-002 & 9820.2909 & 0.5951 \\
            & Text-Embedding-3-Small & 11346.9648 & 0.6876 \\
            & Text-Embedding-3-Large & 19234.9228 & 1.1705 \\ 
            \hdashline
            \multirow{7}{*}{\textbf{\makecell[l]{Fine-tuned \\ code embedding}}} 
            & CodeBERT (110M) & 1734.3351 & 0.0269 \\
            & GraphCodeBERT (110M) & 2613.7416 & 0.0429 \\
            & PLBART (210M) & 2390.2443 & 0.0421 \\
            & CodeT5 (210M) & 4471.2962 & 0.0784 \\
            & UniXCoder (110M) & 2566.1184 & 0.0431 \\
            & CodeT5+ (6B) & 37286.3283 & 0.8294 \\
            & StarCoder (7B) & 41888.5360 & 0.9292 \\ 
            \hdashline
            \multirow{3}{*}{\textbf{\makecell[l]{Zero-shot \\ prompting}}} 
            & Code Llama (7B) & - &  0.2068\\
            & GPT-3.5-Turbo & - & 0.4990 \\ %1.6332
            & GPT-4 & - &  0.5808\\
            \hdashline
            \multirow{3}{*}{\textbf{\makecell[l]{Few-shot \\ prompting}}} 
            & Code Llama (7B) & - &  0.5639\\ 
            & GPT-3.5-Turbo & - & 0.5290 \\ %2.0225
            & GPT-4 & - &  0.6601\\ 
            \hdashline
            \multirow{2}{*}{\textbf{\makecell[l]{Fine-tuning \\ with instruction}}} 
            & Code Llama (7B) & 29206.5457 & 0.5286 \\
            & GPT-3.5-Turbo & 6976.0079 & 0.3156 \\
            \bottomrule
            \bottomrule
        \end{tabular}
    \end{adjustbox}
    % \vspace{-.5cm}
\end{table}

\noindent
\textbf{\emph{\underline{Results.}}}
From Table~\ref{tab:rq4}, we observe that the inference time for the best-performing Compiler-based technique (i.e., TCE$_{Soot}$), the best-performing ML-based technique (i.e., KNN), the best-performing Tree-based NN technique (i.e., ASTNN), and the best-performing LLM-based technique (i.e., UniXCoder) are \SI{2.3537}{s}, \SI{0.0019}{s}, \SI{0.0274}{s}, and \SI{0.0431}{s}, respectively.
On the one hand, compared to traditional Compiler-based techniques, LLM-based techniques have achieved significant improvements in both efficiency and effectiveness. 
On the other hand, while LLM-based techniques may be slightly less efficient than the ML-based and Tree-based NN techniques, their significant effectiveness makes the additional costs associated with LLMs acceptable. 
This indicates that LLM-based techniques provide an excellent balance between cost and effectiveness.

In addition, the pre-trained code embedding strategy significantly outperforms the fine-tuned one in terms of training time.
For example, the pre-trained UniXCoder requires only \SI{809.1785}{s}, whereas the fine-tuned UniXCoder requires \SI{2566.1184}{s}, making the time consumption of the latter three times that of the former.
% It demonstrates that while adopting the fine-tuned code embedding strategy can enhance LLM performance on equivalent mutant detection, it consumes a considerable amount of time during the training phase. 
It demonstrates that the limited computational resource along with the large model size of LLMs is the non-negligible factor that should be carefully considered when adopting the fine-tuned code embedding strategy to enhance LLM performance.
Therefore, in practical usage, there should be a trade-off between efficiency and resource consumption for LLM-based techniques.

% \vspace{-2mm}
\begin{tcolorbox}\textbf{RQ4 Summary:}
The inference time of the best-performing LLM-based technique (\SI{0.0431}{s}) exceeds that of the best-performing Compiler-based technique (\SI{2.3537}{s}) but is marginally longer than that of the best-performing ML-based technique (\SI{0.0019}{s}) and the best-performing Tree-based NN technique (\SI{0.0274}{s}). 
Given the significant effectiveness of LLM-based techniques, a minor increase in inference time is deemed acceptable, highlighting their balance between cost and effectiveness.
\end{tcolorbox}
\section{Discussion}
\label{sec:discussion}

\subsection{Lessons Learnt}

% The study shows that various LLM methods outperform traditional EMD techniques. 
% More figures about the distribution of embeddings can be found on our homepage~\cite{EMD2024}
\textbf{Does the model size affect detection performance?} Our study empirically validated the performance of a series of LLMs with various model sizes. 
The initial assumption of this study was that larger LLMs would have possessed broader prior knowledge and increased learning capacity, thereby enhancing the performance in equivalent mutant detection.
However, our experimental findings indicated that the model size is not the predominant factor influencing LLM performance on equivalent mutant detection.
Conversely, our findings suggest that the data modality and pre-training tasks of LLMs tend to play a more crucial role, a conclusion corroborated by existing studies~\cite{guo2022unixcoder,wang2023codet5+,gao-etal-2021-simcse}.
For instance, UniXCoder surpasses all other studied LLMs, despite its smaller size as shown in RQ1 and RQ2. 
This superiority is likely attributed to UniXCoder leveraging AST to enhance code embeddings with rich syntax and semantic information from source code, achieved through contrastive learning involving three well-designed code-related pre-training tasks.

\begin{figure}[t!]
    \centering
    \includegraphics[width=1\linewidth]{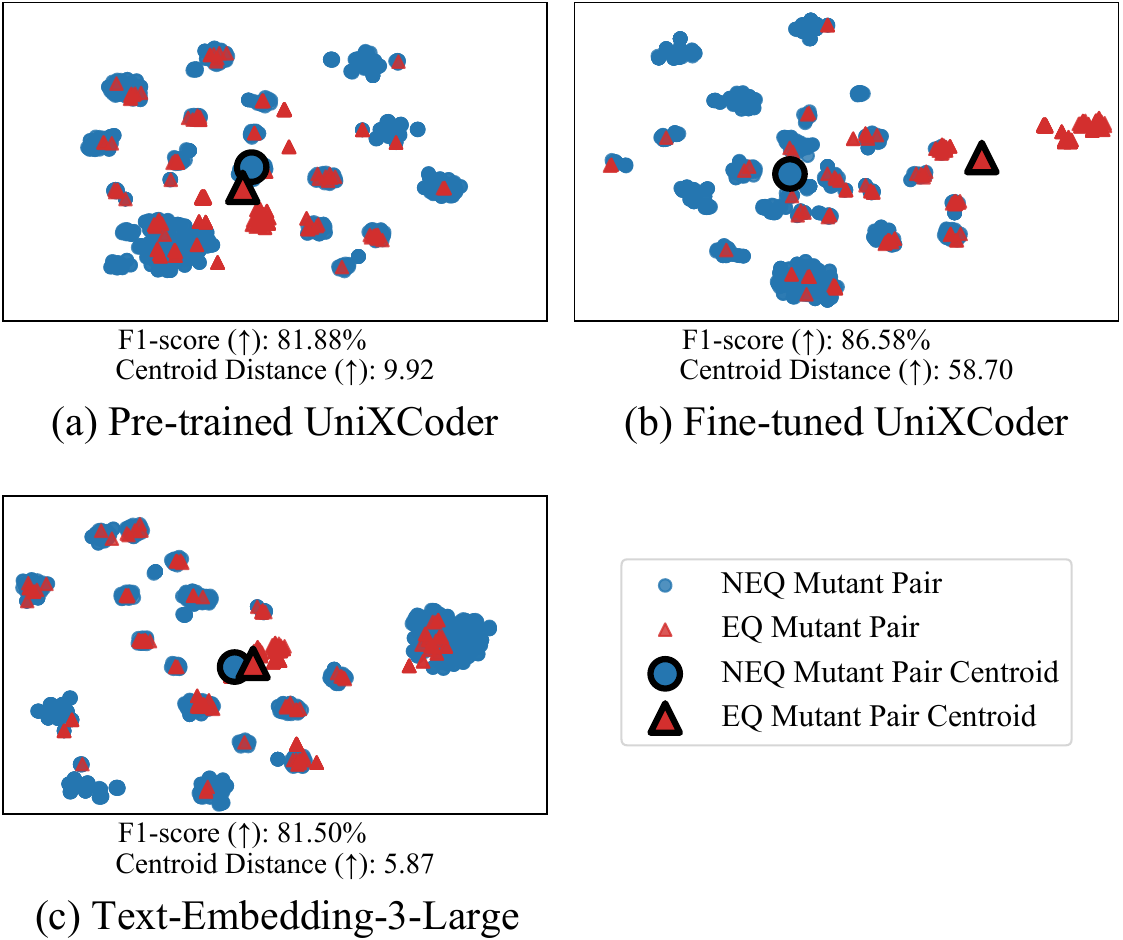}
    % \vspace{-3mm}
    \caption{t-SNE plots showing the embedding of mutant pairs.
    EQ/NEQ represents equivalent/non-equivalent, respectively
    % The red triangle/blue circle indicates equivalent/non-equivalent mutant pair, respectively.
    }
    \label{fig:embedding}
    % \vspace{-.5cm}
\end{figure}
\textbf{Does the embedding quality affect detection performance?}
% The embedding quality crucially affects the classifier's ability to create an effective decision boundary and avoid over-fitting.
% To comprehend the correlation between the quality of the mutant pair embedding and the detection performance, we employ a dimension reduction technique to visualize the embedding within a 2-D space.
% Specifically, following the approach of existing work~\cite{yang2022natural}, we treat each pair of mutant embeddings as a single unit by concatenating them to form a representative embedding. 
% We apply a self-supervised learning method called t-Distributed Stochastic Neighbor Embedding (t-SNE)~\cite{Maaten2008VisualizingDU}, widely used in the deep learning community for evaluating the embedding quality~\cite{DBLP:conf/nips/LiTGLH22, DBLP:journals/jair/BurkartH21}, for mapping the representative embeddings from high-dimensional to 2-dimensional space.
% t-SNE can preserve the local structure among representative embeddings in low-dimensional space, thereby highlighting the quality of the representative embeddings.
% We plot these data points after obtaining the low-dimensional data points of the representative embeddings. 
% The data points are colored according to the labels of the mutant pairs.
% The procedure ultimately produces the embedding distribution in a 2-dimensional scatter plot. 
Existing studies~\cite{DBLP:conf/nips/LiTGLH22, DBLP:journals/jair/BurkartH21} indicate that the embedding quality crucially affects the ability to capture the program semantic feature for identifying an effective decision boundary.
To access the embedding quality of mutant pairs, we employed t-distributed stochastic neighbor embedding (t-SNE)~\cite{Maaten2008VisualizingDU}, which enables us to visually examine the relationship between code embeddings generated by various studied encoder LLMs by projecting them into a 2-dimensional space.
Based on the findings of RQ1 and RQ2, we selected three best-performing LLMs within three categories (i.e., pre-trained code embedding, fine-tuned code embedding, and general text-embedding models). 
Specifically, the selected techniques are pre-trained UniXCoder, fine-tuned UniXCoder, and Text-Embedding-3-Large.
Following existing work~\cite{ahmed2024studying}, we utilized \textit{centroid distance} to measure the separation and quality of code embeddings.
Larger centroid distance values indicate enhanced embedding quality, signifying clearer delineation in the embedding space.

Figure~\ref{fig:embedding} displays the t-SNE plots for all 1,650 mutant pairs in the test set across three studied encoder LLMs.
We find that fine-tuned UniXCoder (58.70) achieves better separation compared to both pre-trained UniXCoder (9.92) and Text-Embedding-3-Large (5.87) in terms of the centroid distance.
The consistent performance of the three LLMs on both the centroid distance (measuring embedding quality) and F1-score (measuring detection performance) metrics suggests a strong correlation between embedding quality and detection performance.

% ==============\tz{the following paragraph need to be rewrote @me}

% Figure~\ref{fig:embedding} demonstrates a slight improvement in the expressiveness of representative embeddings after fine-tuning the encoder. 
% When comparing pre-trained UniXCoder and fine-tuned UniXCoder, the embeddings of equivalent mutant pairs are better aggregated. 
% Some incorrect clustered representative embeddings have been dispersed. 
% However, there are still embeddings of equivalent mutant pairs mixed in with clusters of non-equivalent pairs.
% Conversely, the distance between equivalent and non-equivalent embedding clusters in UniXCoder is larger than in Text-Embedding-Ada-002 and Text-Embedding-3-Large. 
% This reveals that the embeddings generated by the two GPT-related models lack strong cohesion within the equivalent cluster, and the difference between varying cluster types is not adequately pronounced.
% These findings confirm that: (1) The distribution of mutant pair embeddings relatively aligns well with the results from RQ1 and RQ2, and (2) the generated mutant pair embeddings fairly support the EMD classifier in identifying an optimal decision boundary, thus avoiding over-fitting.

\subsection{Future Work}
\label{subsec:future}
% This work also provides several promising future directions:
% In this paper, we provide the first comprehensive evaluation of LLMs for equivalent mutant detection, yielding numerous key findings.
% Despite having conducted extensive experimental evaluations, there are some aspects in our study that can be further extended.

\textbf{Across Different Programming Languages.}
Due to the limited computational resources and time cost, we selected the popular Java as the representative programming language for the evaluation.
In the future, we will extend our experimental evaluation in other languages to comprehensively explore the performance of LLMs in equivalent mutant detection. 
Moreover, we plan to analyze the performance of LLMs in detecting equivalent mutants across different programming languages. 
This investigation will facilitate an in-depth analysis of LLMs' understanding capability of various syntaxes and structures across diverse programming paradigms.

\textbf{Chain-of-Thought Prompting.} 
In our study, we solely employed the typical prompting techniques (i.e., zero-shot prompting and few-shot prompting).
Recently, Chain-of-Thought (CoT)~\cite{kojima2022large,tian2023test} prompting technique has been proposed, facilitating LLMs for tackling complex problems (e.g., mathematical reasoning and code generation), through an intermediate reasoning process to derive final solutions.
Several studies have confirmed the effectiveness of CoT prompting in enhancing LLM performance across complex reasoning benchmarks~\cite{wei2022chain,jiang2023self}.
Hence, we can further investigate the role of CoT prompting in the task of equivalent mutant detection. 

\textbf{Equivalent Mutant Avoidance.} 
Rather than detection, some research focused on avoiding the generation of equivalent mutants~\cite{madeyski2013overcoming}. 
These avoidance techniques often involve meticulous construction of mutants employing program dependence analysis or higher-order mutation operators to reduce the number of equivalent mutants~\cite{harman2001relationship,jia2009higher,oh2021effectively}.
Equivalent mutant detection techniques and equivalent mutant avoidance techniques are orthogonal to a large extent. 
% Equivalent mutant avoidance techniques reduces the number of equivalent mutants before the mutant generation, while equivalent mutant detection techniques can still be applied to further identify and reduce the remaining equivalent mutants after the mutant generation.
Future work could explore the synergy of these two categories to further enhance the mutation testing process.

\textbf{Duplicated Mutant Detection.} 
% Equivalent mutants have proven to be a significant challenge in mutation testing.
A related issue is the problem of mutant duplication.
Duplicated mutants refer to mutants that are semantically equivalent to some other mutants, although both duplicated mutants may be semantically different from the original program. 
Duplicated mutants are also a challenge for mutation testing as they may inflate the mutant-killing effectiveness of a test suite.
% Specifically, assuming all other factors remain constant, a test case that kills multiple duplicated mutants holds no superior effectiveness than another test case that kills only a single non-duplicated mutant.
~\citet{kintis2017detecting} demonstrated that the equivalent mutant detection technique (i.e., TCE) can directly detect these duplicated mutants.
In the future, we will delve into exploring the performance of LLMs in duplicated mutant detection.

\subsection{Threats to Validity}
\label{subsec:threats}
% \shl{temp=0 decrease randomness}
\paragraph{External threat} This threat mainly lies in the equivalent mutant dataset used in our study. 
We only focus on programs written in Java, thus our results may not be generalized to other languages. 
In future work, we plan to extend our study framework to investigate LLM performance across diverse programming languages in equivalent mutant detection. 

\paragraph{Construct threat} Three related threats are summarized.
First, the construction of training and test datasets may introduce potential bias resulting from the adopted strategy.
However, the stratified sampling strategy and the setting of 50\% are commonly used.
Second, we acknowledge that the EQ/NEQ ratio we used is not perfectly realistic.
Nevertheless, our used ratio (17.80\%) better reflects practical scenarios compared to the 50.00\% ratio typically used in existing studies~\cite{brito2020preliminary,peacock2021automatic,ma2023scope}. 
In future work, we will conduct a more comprehensive study to investigate the LLM sensitivity across various ratios in the more practical benchmarks.
Last, due to the limited computational resources and cost, we did not run our studied EMD techniques multiple times to mitigate potential variance and randomness. 
Future work is encouraged to repeat the experiment multiple times and report the average results.
% \wang{I copy this from a fuzzing paper, needs to be updated accordingly.. maybe construct threat can be from (1) the preparation of training and testing dataset and (2) the randomness of experiment since we did not run LLMs in multiple rounds due to the computation resource.}
% To address this issue, on one hand, we have taken care to select 9 representative LLMs as our base models and 10 EMD techniques as the baselines. 
% On the other hand, considering the inherent sensitivity of LLMs (i.e., Code Llama, GPT-3.5-Turbo, and GPT-4) to prompts, we set the decoding temperature to 0 to help further reduce randomness.

\paragraph{Internal threat} This threat mostly lies in the implementations of each studied EMD technique.
To mitigate this threat, we implemented EMD techniques based on the open-source tools of each paper, and three authors have carefully reviewed the source code.

\section{Conclusion}
\label{sec:conclusion}
This work conducts an empirical study to extensively investigate the effectiveness and efficiency of LLMs for equivalent mutant detection.
Specifically, we assess the performance of ten studied LLMs in comparison to ten existing EMD techniques, examine the various strategies of LLMs, evaluate the orthogonality between EMD techniques, and measure the time overhead of training and inference.
The key findings highlight that LLM-based techniques significantly surpass all baselines, with the fine-tuned code embedding strategy being the most effective. 
Moreover, LLM-based techniques strike an excellent balance between cost and effectiveness.
% Our work also paves the way for promising future research including the study of equivalent mutant detection across different languages, understanding the role of chain-of-thought prompting, and the application of LLMs to duplicated mutant detection.
Our work also paves the way for promising future research such as the study of cross-language equivalent mutant detection, chain-of-thought prompting, combined effects with equivalent mutant avoidance techniques, and LLM application in duplicated mutant detection.

\section{Data Availability}
\label{sec:data_availability}
We released all the experimental data and source code on the project homepage for replication, future research, and practical use~\cite{EMD2024}.

\begin{acks}
This work was supported by National Natural Science Foundation of China (Grant Nos. 62322208, 62232001), CCF Young Elite Scientists Sponsorship Program (by CAST), JSPS for the KAKENHI grants (JP21H04877, JP22K18630), Bilateral Program grant JPJSBP120239929, and the Inamori Research Institute for Science for supporting Yasutaka Kamei via the InaRIS Fellowship. 
\end{acks}

\balance
\bibliographystyle{ACM-Reference-Format}
\bibliography{reference}

\end{document}